# Blockchain for Cities—A Systematic Literature Review

## CHARLES SHEN [ORCID]1 AND FENIOSKY PENA-MORA2
[1]Advanced Construction and Information Technology Laboratory, Columbia University, New York, NY 10027, USA
[2]Center for Buildings, Infrastructure and Public Space, Columbia University, New York, NY 10027, USA

Corresponding author: Charles Shen (charles.shen@columbia.edu)

This work was supported by the Kong Family and Zhao Family Gifts.

**ABSTRACT** Blockchain is considered one of the most disruptive technologies of our time. Numerous cities around the world are launching blockchain initiatives as part of the overall efforts toward shaping the urban future. However, the infancy stage of the blockchain industry leads to a severe gap between the knowledge we have and the actions urban policy makers are taking. This paper is an effort to narrow this rift. We provide a systematic literature review on concrete blockchain use cases proposed by the research community. At the macro-level, we discuss and organize use cases from 159 selected papers into nine sectors recognized as crucial for sustainable and smart urban future. At the micro-level, we identify a component-based framework and analyze the design and prototypes of blockchain systems studied in a subset of 71 papers. The high-level use case review allows us to illustrate the relationship between them and the four pillars of urban sustainability: social, economic, environmental, and governmental. The system level analysis helps us highlight interesting inconsistencies between well-known blockchain applicability decision rules and the approaches taken by the literature. We also offer two classification methodologies for blockchain use cases and elaborate on how they can be applied to stimulate cross-sector insights in the blockchain knowledge domain.

## I. INTRODUCTION

Cities are facing tremendous pressures from challenges associated with rapid urbanization. The United Nations estimates that 4.2 billion or 55% of the worlds' population lives in urban areas in 2018 and a 25 billion or 13% will be added by 2050 [1]. City growth gives rise to not only population explosion, but also severe issues such as traffic congestion, pollution, non-renewable resource depletion, and increasing social inequality [2]–[4]. These urban problems do not respect borders of the nations or limits between industrial domains [5]. Great responsibilities on solving them lie at the city level, where conflicts regarding economic, social and environmental development are often managed. Some even say that "Mayors rule the world" [6].

For the past decades, numerous urban sustainability and smart cities frameworks have been proposed [7]–[10]. They provide tools to help urban policy makers make decisions, take actions, and assess the cities' progress towards a more sustainable future [11].

Recently, researchers started to advocate the notion of "blockchain cities" [5] as the next wave in transforming the urban context to meet the urbanization challenges. In this regard, blockchain may be compared to a General Purpose Technology [12] that is "complementary to human and organizational capital and whose usage is shaped by political choice and by the urban ecosystem of the citizens, technology vendors and local authorities, depending on the city's needs and habits" [10]. Many believe that blockchain is poised to play an important role in the sustainable development of the global economy [13], improving people's quality of life and ultimately bringing fundamental changes to the world we live [14]. A World Economic Forum report estimates that 10% of global GDP will be stored on blockchain technology by year 2027 [15].

Blockchain features a decentralized shared database that provides transparency and immutability of transaction records. Initially implemented by the well-known cryptocurrency Bitcoin [16], it later evolves to offer a trust-free platform for execution of arbitrary business logic through many alternative blockchain platforms, for instance, Ethereum [17] and Hyperledger Fabric [18]. These continuously evolving blockchain technologies are increasingly considered a disruptive force to virtually every sector of the society [19]–[22].

Many cities around the world have reported blockchain-related initiatives, such as those in Australia, China, Denmark, United Arab Emirates, Estonia, Georgia, Ghana, Honduras, Malta, Russia, Sweden, Singapore, Spain, Switzerland, United Kingdom, Ukraine, United States (US) [23]–[27]. Cities and states set up various goals and employ many approaches in the race to lead the blockchain wave. For example, Dubai is building a single software platform through which city public sector can launch blockchain projects, as part of the ambition to become paperless by 2020 [28]; in contrast, Illinois takes a more experimental approach, launching multiple separate blockchain pilots across different industrial sectors including governance, education, health-care and energy, each selecting their own blockchain platform as appropriate [28]. Zug is developing itself to be a "crypto valley" through establishing a crypto-friendly business ecosystem [29]. New York City announced plans to launch the Blockchain Resource Center as a hub for the city's blockchain industry and to convene both government and citizen stakeholders in developing a regulatory environment that stimulates the overall blockchain industry [30].

Despite all the ongoing blockchain efforts, many also believe that our current understanding of blockchain is premature and there is a lack of knowledge on where blockchain technology can provide mentionable societal effects [31]. Sometimes the field is even described as "an innovative technology searching for use cases" because it is largely unknown how blockchain could be incorporated into existing digital services, processes and infrastructures [22]. In a testimony to the US congress, US Department of Homeland Security's Science and Technology Division Director Douglas Maughan also pointed out specific concerns in the blockchain space for the asymmetries between knowledge and action [32]. Biased use of the buzzword in fragmented or superficial ways will lead to more confusion than clarity. Falling into the tendency to believe that innovative technologies like blockchain can automatically transform the ecosystem around us will actually hinder the achievement of the technology's real potential.

Under this mixed backdrop, this paper attempts to advance the understanding towards how blockchain can fit in the next level of urban development initiatives, by combining foundational frameworks on sustainable and smart cities with blockchain domain knowledge accumulated by the research community. Through helping city policy makers, industrial practitioners and all stakeholders better understand blockchain use cases in cities, we hope to facilitate decision makers in planning of blockchain strategy and drive actions in the most pertinent industrial domains that contribute to meet the urban growth challenges. Our work also serves to reinforce the notion that blockchain technology by itself will not transform the city; instead the change requires a political understanding of technologies, a process approach with focus on all aspects of public values.

Our research employs a pragmatic methodology to provide a comprehensive literature review on the blockchain use cases studied in academic journals and conference proceedings. Given the enormous number of papers about blockchain, we confine our work to the portion of papers that focus on concrete use cases and with extensive system coverage. Our main contributions are in the following areas:

- We provide an application-oriented use case review of 159 selected papers, and organize them based on a list of 9 industrial sectors central to 16 major global urban sustainability and smart city reference frameworks as identified in [7]. Our review shows how blockchain-enabled innovations are changing the urban systems and the ramification of these changes for different sectors of the society. We also illustrate how our application-oriented review can be used to evaluate the blockchain application efforts for urban sustainability goals.
- We dive deeper into a subset of 71 papers to examine more details about their design and implementation choices. To facilitate the analysis, we propose a component-based general analysis framework for blockchain use cases that covers both the external and internal factors of the blockchain system. We further show that the component-based analysis can help identify gaps in actual blockchain use cases versus common blockchain applicability criteria.
- We propose two classification methods for blockchain use cases, one role-based and the other business model based. We also demonstrate how these taxonomies can provide insights for cross-sector blockchain application design and analysis.

The structure of this paper is as follows: we present background knowledge on blockchain technology in Section II and introduce our research methodology in Section III. Section IV provides related work. The next two sections, Section V and Section VI contain our main application-oriented review and components-based analysis of the blockchain use cases, respectively. They are followed by Section VII which conducts a further discussion on the analysis results. Lastly, Section VIII concludes the paper, summarizes its contributions and limitations, and suggests future work.

## II. OVERVIEW OF BLOCKCHAIN

Table 1 lists three representative blockchain technologies and their main characteristics, which will be elaborated below to give an overview of the topic.

### A. BITCOIN CRYPTO CURRENCY BLOCKCHAIN

The blockchain concept is known to originate from a paper on Bitcoin [16] published in 2008 by someone in the name of Satoshi Nakamoto. Bitcoin is a decentralized crypto currency and remains the most important blockchain application today. It is believed that the inventor created Bitcoin to offer an alternative to the central-bank controlled monetary system, which many people consider as a cause of the global economic crisis around 2008.

**TABLE 1.** Representative blockchain technologies.

| Blockchain Technologies | Bitcoin [16] | Ethereum [17] | Hyperledger Fabric [18] |
|---|---|---|---|
| Distributed Consensus | Proof-of-Work mechanism | In transition from Proof-of-Work to Proof-of-Stake mechanism | Byzantine Fault Tolerance and other mechanisms |
| Crypto Token | Bitcoin native token critical to sustain the crypto currency ecosystem | Ether native token critical to sustain the decentralized computing platform | Native token not applicable, but application level tokens are possible |
| Business Logic Support | Very limited scripting | Fully programmable smart contract | Fully programmable smart contract |
| Participation Model | Permissionless, anyone can join | Permissionless, anyone can join | Permissioned, only authorized party can join |

A typical blockchain consists of a peer-to-peer network of computer nodes that maintain a decentralized shared database of records. In the original Bitcoin blockchain, the records contain transfer transactions of Bitcoin crypto currency between participating parties. Each party in the transaction has a Public Key Infrastructure (PKI) private key and public key pair. The hash value of the public key is used as the party's identity or transaction address. Transaction parties sign the transactions using their private key, which could later be verified by other parties using the signer's public key. The transactions are broadcast to all peer nodes in the network. Using a distributed consensus mechanism, the peer nodes agree on what transactions are valid and the sequence of those transactions that take place. These transactions are placed into a data structure called "block" and committed to the shared database to form a linked chain, hence the name "blockchain". Each block in the blockchain has its own timestamp and a cryptographic hash that connects it to the prior block. Blocks can only be appended, not deleted. The outcome is a shared database with an ever-growing list of records that are immutable and irreversible; tampering of any block information can be detected by peer nodes on the blockchain.

### 1) PROOF-OF-WORK DISTRIBUTED CONSENSUS

Distributed consensus mechanism is critical for blockchain since it determines which block can be accepted and inserted to the chain. This is akin to agreeing on distributed power allocation because the node authoring the accepted block (hereafter referring to as the official validator) is able to change the state of the database shared by every other peer. In order to secure this process, the power allocation has to be associated with some cost and resources to prevent abuse. The solution employed by the original Bitcoin blockchain is called proof-of-work, in which nodes have to compete by calculating a cryptographically sophisticated puzzle. The characteristics of this puzzle ensures three properties: a node has to invest corresponding amount of computing power to complete it; the next node to successfully solve the puzzle is random; and a node's claim on finding the answer of the puzzle can be easily verified by any other peer nodes. One additional issue is, however, malicious nodes controlled by an attacker could also be randomly selected as official validator as long as they follow the same process. Once chosen, a malicious node could still try to inject blocks of false transaction records into the blockchain. Therefore, there is a follow-up implicit consensus step after a peer node receives the block proposed by the official validator. In this step, the peer nodes can verify the transactions in the received new block, and if any anomaly is detected in it (such as inconsistency of the linked hash values, or mismatched transaction signature and identity), they can keep the prior state of the blockchain without accepting the new block. Otherwise if everything goes well, the node confirms the new block and accepts the updated blockchain. The likelihood of a block being rejected diminishes exponentially with the number of acceptance confirmations it receives from different nodes. After a certain number (e.g., 6 in the case of Bitcoin) of confirmations, the block is considered permanently committed to the blockchain.

### 2) CRYPTO TOKEN ECONOMICS

In addition to using distributed consensus mechanism to prevent nodes from misbehaving, the blockchain can also use crypto token asset to proactively incentivize desired node behavior. In particular, the official validator is rewarded some crypto tokens for its efforts in validating and packaging the new block of transactions whenever it introduces a valid block that gets accepted into the blockchain. This rewarded crypto token can be created (mined) when a new block is inserted into the blockchain (the process is called crypto token mining), or it can be paid by the initiators of the transactions in the block as a service fee. If the official validator tries to introduce a block with invalid transaction information, however, that block could be rejected by peer nodes from the blockchain. The official validator then lose the crypto token rewards associated with that block.

In the Bitcoin blockchain, the corresponding crypto token is Bitcoin. Bitcoin's value is established upon its utility as a currency payment method and its expected appreciation of the future values, leading to a liquidity market between the crypto token and fiat currency. This essentially creates a crypto token economic model around the Bitcoin.

### B. FROM CRYPTO CURRENCY TO EVERYTHING

The notion of blockchain soon expands from crypto currencies to general purpose business areas.

### 1) ETHEREUM AND SMART CONTRACTS

Ethereum [17] represents the next major leap after Bitcoin in the blockchain space with its full support for smart contracts. A contract is a fundamental piece of our market economy and defines relationships among both businesses and individuals. The smart contract concept is originally proposed by Szabo in 1994, who states that ''The general objectives of smart contract design are to satisfy common contractual conditions, minimize exceptions both malicious and accidental, and minimize the need for trusted intermediaries [33]''. To achieve these objectives, smart contract clauses expressing business logic can be encoded in computer programs and automatically executed with computer-based systems. The vending machine carries a primitive form of smart contracts by implementing a simple transaction of accepting coins and returning corresponding goods.

For generic smart contracts, the blockchain is arguably a perfect infrastructure as it provides a transparent and traceable platform that allows parties to perform trust-free transactions with each other without intermediaries. In this paradigm, smart contracts are computer programs written in a language supported by the underlying blockchain platform. The programs are automatically executed according to the designated triggering conditions in the contracts. These conditions can be outcomes of transactions or interactions with other smart contracts. The conditions can also be triggered by external events. Since there is no native way for smart contracts to directly interact with external systems, Oracle services act as the bridge to provide a single truth view of the external system for smart contracts.

The critical characteristics of blockchain-based smart contracts compared with traditional contracts can be summarized in three aspects. First, smart contracts running on a blockchain are entirely managed by computer code and not subject to control of any central entity. Second, the only way to modify a deployed smart contract is to create a new one under the consent of all involved parties. The old one cannot be simply retracted. Third, it is cost-effective to establish multi-party agreements requiring multiple conditions, resulting in great flexibility.

Smart contract can also be modeled as a state machine [34]. After its execution the states across network nodes will be consistently updated; blockchain's consensus process achieves this update and therefore some people compare blockchain as the operating system for smart contracts [35].

Ethereum is the most well-known blockchain platform for running smart contracts. It has an integrated Turing-complete computer language by design, which means it can support any kind of general purpose programs. Execution of the code is through a virtual machine and costs ''gas'' fees paid through its native Ether crypto token. The charge of gas fees is to prevent the computer system from being abused and entering dead loop. In contrast, the original Bitcoin blockchain offers very limited scripting capability and is only able to support rudimentary smart contracts, if at all. But that could be intentional by the Bitcoin designers since it is reasonable for a crypto currency application not to allow arbitrary programmable manipulation in order to mitigate security risks.

### 2) GENERIC CRYPTO TOKEN SYSTEMS

As smart contracts widely open up the spectrum of possible blockchain applications, the meaning of crypto tokens in blockchain also evolved from a digital currency to representing any tradable asset, from fungible goods such as movie tickets, loyalty points, company shares, to non-fungible things such as software license. Use of these tokens makes programming asset exchange possible and executing the business logic easier. For that reason, blockchains such as Ethereum provide standard mechanisms to facilitate token issuance, distribution and exchange, including the ERC-20 tokens [36] for fungible and the ERC721 token [37] for non-fungible and indivisible assets.

The expanded role of crypto tokens comes with appropriate token economic models. The two common models are utility and security. In the utility model, tokens provide utility values. For example, the Ether token serves as the payment method for transaction fees on the Ethereum platform at the infrastructure level, and Bitcoin tokens can be used as a payment method for assets exchange at the application level. In the security model, crypto tokens function like securities. At the infrastructure level, they may entitle their holders to the mining rights of the blockchain, or voting rights on the directions of the platform development, or profits sharing with the platform (e.g., from transaction fees). At the application level, crypto tokens held in a decentralized autonomous organization can represent governance rights in the organization.

Regardless of the utility or security model, the crypto token's function as incentives is still carried over in many blockchain applications. Besides providing infrastructure level incentives to secure the network as in the original Bitcoin blockchain, these incentives are also applied at the application level, e.g., a use case promoting sustainability may reward participants for their use of environmental friendly transportation methods.

### 3) PROOF-OF-STAKE DISTRIBUTED CONSENSUS

In the distributed consensus space, there have been a lot of developments addressing a common complaint about Bitcoin's proof-of-work mechanism, specifically, it consumes too much energy because of the intensive computational requirements. A popular alternative proposed is called proof-of-stake. It chooses the node to be the official validator based on the proportional stake of the network value that each node holds, and therefore eliminating the computational cost. But the proof-of-stake mechanism has its own problems, notably the ''Nothing at stake'' issue which in the event of conflicts could prevent a blockchain from convergence and result in forked chains; and the ''Long-range attack'' issue where the longest fork of the chain may be replaced by a chain reconstructed from the genesis block. There are various efforts trying to address these shortcomings. For example, the delegated proof-of-stake approach lets nodes elect delegate nodes to serve as designated validators for new blocks, and these nodes can be out-voted if they do not behave properly. While delegated proof-of-stake is more resilient to the problems with the original proof-of-stake mechanism, it also reduces the system's degree of decentralization.

### C. FROM PERMISSIONLESS TO PERMISSIONED

In the Bitcoin blockchain model, any node can join and participate in the peer-to-peer network, so it is called a permissionless blockchain. It is similar to the Internet model, where virtually any node can connect and become part of it. The Ethereum blockchain also belongs to the permissionless blockchain category. Afterwards, the blockchain community introduced the permissioned blockchain model. The proposition for a permissioned blockchain comes from a different mentality and is more suitable for an industrial consortium or enterprise environment. It enables participants to be authorized before they can join the network and be assigned appropriate functional privileges in the network matching their respective identities. If every participation and access of the blockchain needs to be controlled, it is also called a private blockchain. A permissioned blockchain contains a logically centralized trusted identity management system that issues cryptographic certificates to qualified participants, and a distributed database under a decentralized administration, providing improved transparency and auditability across the involved parties than in traditional distributed databases.

#### 1) BYZANTINE FAULT TOLERANCE DISTRIBUTED CONSENSUS

The consensus mechanism in permissionless blockchains are usually competitive in nature. That is because the nodes in those blockchains do not trust each other, and having them put up some stake (e.g., computer power, economic cost) in order to participate in the consensus outcome protects the security of the network. In a permissioned blockchain, the official validators of new blocks are known. This opens up a broad collection of distributed consensus protocol possibilities. A popular family is based on state-machine replication with byzantine fault tolerance [38], which has the capability to function successfully in the presence of certain number of malicious or faulty nodes. In a typical practical byzantine fault tolerance environment, nodes are divided into clients and validators. Validators manage public key infrastructure identity and certificate authority. The clients send their transactions to a primary validator, which in turn broadcast the information to other validators. Those validators process the transaction and send response back to the original client. The client collects at least one third of the same results from all the validators to confirm the transaction. This mechanism can achieve a much higher performance, a throughput in the order of tens of thousands of transactions per second compared to 7 transactions per second in the original Bitcoin blockchain. Nevertheless, the scalability of state-machine replication with practical byzantine fault tolerance has not been proven [34], and it is typically suitable only for systems with a relatively small number of nodes [20].

Hyperledger Fabric [18] is a well-known permissioned blockchain and provides multiple algorithm options for the consensus process, including byzantine fault tolerance algorithms.

#### 2) SMART CONTRACTS

In terms of smart contracts, a permissioned blockchain may also provide full smart contract capabilities, such as the Hyperledger Fabric blockchain or they may provide only limited capabilities, such as the MultiChain platform [39].

#### 3) CRYPTO TOKEN MODELS

In permissionless blockchains, the crypto token models create opportunities for economic alignment, shared interest, and coordination between distributed and trustless individuals [35]. Token holders on the blockchain naturally have a vested interest in the success of the specific crypto token and its underlying blockchain infrastructure that supports its utility or security value.

In contrast, permissioned blockchains have centralized control on node participation. They generally do not need the type of crypto tokens serving as incentives to sustain the blockchain infrastructure, even though application-specific tokens can still be applicable.

### D. BLOCKCHAIN SECURITY

A peer-to-peer blockchain network entails important security risks especially when it is permissionless where anyone can join. Some of the most important vulnerabilities of blockchains are the following:

Double spending: in a Bitcoin type of crypto currency payment network, a malicious party may attempt to pay the same units of crypto currency simultaneously to two different parties. This is called a double-spending attack. In general, the network needs to make sure that once one of those transactions is accepted, the other one will be rejected in order to prevent double spending.

Sybil attack [40]: a malicious party could create many nodes all under his own control to increase his chance of being selected as the official validator and control the blockchain. This problem is why resources are required to participate in the validator selection and crypto token incentives are offered to encourage proper behavior.

51% attack: an attacker could compromise the blockchain by trying to obtain overwhelming resources. In the case of proof-of-work, owning 51% of the computing power would control more than half of the block validator opportunity and also significantly improve the success possibility of other attacks such as double spending. A sufficiently large network deters the 51% attack by the enormous amount of resources required to launch it.

Denial-of-Service attack: a malicious node could refuse to add a valid transaction into the blockchain, essentially denying service to the particular transaction party. This risk is mitigated by blockchain's de-centralized network architecture. Since the transaction information is broadcast to all nodes, it is hopeful that at least some nodes will process it if the transaction is valid.

## III. METHODOLOGY

### A. BLOCKCHAIN USE CASE CATEGORIES UNDER SUSTAINABLE AND SMART CITY CONTEXT

Due to the infancy stage of the technology, strategic planning for blockchain applications in cities is still a largely unknown field. Meanwhile, blockchain technology as a digital innovation is conceptually related to the key information and communications technologies underlying smart cities. We therefore turn to the existing smart cities and its well-known predecessor, sustainable cities framework as references. Those frameworks have become the de facto goals for cities around the world for decades, providing a plausible anchor point for organizing our research on blockchain for cities. Smart cities are commonly assessed based on prior experiences on sustainability and quality of life, with a significant addition of modern technological components [41], but there are important differences between sustainable and smart city indicators [7].

In order to create a meaningful discussion of blockchain for cities with both smartness and sustainability goals in mind, we adopt the essential city sectors for sustainable and smart cities identified by [7] as a result of its thorough examination on 16 of the most well-established sets of assessment frameworks, 8 on smart cities and 8 on sustainable cities. These categories include "Governance and citizen engagement", "Education, culture, science and innovation", "Well-being, health and safety", "Economy", "Transportation", "Energy", "Water and waste management", "Built environment", "Natural environment", and "Information and Communications Technology (ICT)". The only exception we made is the exclusion of the ICT sector, which is not to undermine its importance, but is a trade-off from the scope of this paper which focuses on blockchain technology applications in urban sectors other than the digital technology itself.

### B. RESEARCH QUESTIONS

With the industrial sectors defined, we formulate the following research questions for our review:

- What are the blockchain use cases studied by the research community in the sectors central to smart and sustainable cities?
- What could be a unified framework to examine those use cases regardless of their sectors?
- How do we evaluate the impact of the use cases on urban sustainability?
- How do we evaluate the blockchain applicability of the found use cases?

- What appropriate taxonomies of blockchain applications can be derived to facilitate cross-sector use case analysis?

To answer the above research questions, we refer to best practices for a systematic review [42]–[44] and follow a standard protocol for selecting literature to be included in the review.

### C. DATA SOURCES

The sources of search include both the main major-focused databases ACM, ASCE, IEEE, e-Government Research Library (EGRLv13.5) [45] and multidisciplinary databases JSTOR, ScienceDirect, Scopus, Springer, Web of Science, Wiley. Since this particular study is interested in scientific knowledge on the blockchain application for cities, we focus on literature in academic journal and conference proceedings, which helps to ensure quality [31], [43], [46].

### D. SEARCH PROCESS

We posed search term hypothesis and conducted test search to establish an overview of the topic as the foundation. Initially we used a combination of keywords including *blockchain* and *city* or *urban*, but found those search terms too limited. Blockchain application has been proposed on virtually every aspects, many of them touching certain city sectors but the papers do not necessarily state city or urban explicitly. Therefore, we eventually settled on finding all papers only with the term *blockchain*, as in most of the related work [47]–[49]. This process results an initial number of 3827 papers. It is worth noting that this number is not much different from the total number of papers a related work [31] found back in January 2017 from a similar list of sources, even though we would expect a number much larger given the highly intensified attention to this topic through 2017 and 2018. This could be caused by the fact that [31] used both *blockchain* and the non-concatenated word *block chain* as search terms. During our test search, we found that the non-concatenated term could produce large number of false positives, basically papers referring to block chain notions in scientific fields very different from the blockchain we are concerned about. Even for a small number of papers where they do refer to the blockchain of our discussion, they often also include the concatenated version of the term *blockchain* either in full paper body or cited references. So we choose to use the *blockchain* term as with most of the other related work.

### E. SCREENING PROCESS

We then examined title, abstract and keywords, and when unsure, the full text body of the retained papers to select the ones that are relevant to our research questions. Given the huge number of papers in question, we have to properly confine our scope and at the same time avoid sacrificing our research goals. Our scope limit is defines as follows:

We select papers that present sufficient depth of system design or prototype evaluations on concrete use cases in the 9 sectors we articulated in Section III-A: "Governance and citizen engagement", "Education, Culture, Science and Innovation", "Well-being, health and safety", "Economy", "Transportation","Energy", "Water and waste management", "Built environment", and "Natural environment". We explicitly excluded papers in the following categories: conceptual discussions of blockchain and its trends, papers focusing on aspects that are more generic and agnostic to sectors, such as improving blockchain technology itself, ICT and identity management, and other technical papers proposing algorithms without an emphasis on explicit use cases in the sectors we identified. We also excluded cryptocurrency related use cases as that has been the focus of most of the prior reviews and also is more concerned at the central government level. However, we do include cases that use cryptocurrency payments as part of the mechanism if they fit our other selection criteria. The above process leads to a total number of 159 non-duplicated papers across all the 9 sectors for our first part, application-oriented use case review. For the second part of our review which dives deeper into system component analysis, we further nail down to a subset of papers that provide more details concerning our component-based framework. That process renders 71 of the 159 papers across 8 of the 9 sectors.

## IV. RELATED WORK

Reviews of blockchain research in the recent years show that the majority of scholarly work has focused on improvements and challenges of current protocols, primarily for cryptocurrencies in general and for Bitcoin in particular [14], [47], [50], [51]. Little is on research that delves into purported disruptive potential of blockchain [52]. While research on some areas especially cryptocurrencies and payments are well developed, comprehensive understanding regarding application and use cases is generally missing [31].

Table 2 compares our work with the related systematic literature review work we found. In addition to a more updated list of papers examined, the focus of our work differs from all of them. Our work explicitly excluded pure cryptocurrency systems and instead investigates general purpose blockchain use cases, while [47], [48], [51] are all primarily focused on Bitcoin and cryptocurrency literature. Reference [53] is not a general purpose blockchain use case review but particularly discussing the impact of blockchain characteristics on service systems. Reference [49] is a bibliometric study reporting the number of blockchain papers in the surveyed set in four categories: Internet of Things (IoT), Smart Contracts, E-governance, and Others, with brief explanation on each category. Lastly, [31] proposes a conceptual framework for blockchain research adapted from recognized social media research agenda. It features an intersection of activities between user and blockchain developers at different levels of analysis, and serves to stimulate multidisciplinary research approaches on blockchain. The component-based framework proposed and followed in this paper, however, takes on a very different perspective that geared toward cross-sector analysis for system design and implementations.

There are much more related work that center on specific domains with various levels of depth and may or may not adopt a systematic literature review approach, for instance, on finance and cryptocurrencies [54]–[56], governance [24], education [57], energy [58], IoT [59], healthcare [60]. Our work differs from them in that we stress a use case study across all sectors in the smart and sustainable city context, and apply the proposed component-based analysis to drive cross-sector insights. To the authors' knowledge, we are the first to provide a systematic blockchain use case review with this methodology.

## V. APPLICATION-ORIENTED USE CASE REVIEW BY SECTORS

The 159 papers selected for our application-oriented review in all sectors are shown in Table 3. Even though one use case could involve aspects from multiple sectors, we place each use case into only one primary sector to facilitate the discussion. We will then relate use cases from various different sectors in later sections.

### A. GOVERNANCE AND CITIZEN ENGAGEMENT

Digital governance contributes to the crucial sustainable development agenda [219], [220], for example, in reducing corruption, lowering administrative costs, insuring document integrity, connecting donors and disadvantaged groups like refugees and displaced people [221], [222].

To look at how blockchain as a digital technology is likely to have a significant impact on city governance, it is helpful to start with the four ideal-typical conceptualizations of smart city governance identified by [223]. They include: (1) government of a smart city, (2) smart decision-making, (3) smart administration and (4) smart urban collaboration. These four models represent an increasing level of progressiveness from more conservative to more radical. The "governance of a smart city" model concerns about setting up the right policy choice and effectively implementing the initiatives under the traditional governmental structures. It is the most basic and common model. The "smart decision-making" model entails restructuring of the decision-making process. An example of this model is urban decision-making leveraging big-data collected from IoT sensor networks. This model involves a certain level of transformation in the process but not at the government organization itself. The "smart administration" model requires using sophisticated Information Technology (IT) to interconnect information, processes, institutions, and physical infrastructure to better serve citizens and communities [224]. It thus leads to re-structuring of existing government organization to integrate traditional functions of government and business [225]. Lastly, the "smart urban collaboration" model is the most transformative which requires transformation at both internal and external of the government organizations. It emphasizes truly citizen-centric operations and services based on collaboration across departments and communities [226], a technology-enabled community-based model of governance [227] and a proactive, open-minded governance structure where all actors work together to maximize the urban sustainability and minimizes negative externalities [228].

**TABLE 2.** Related work on systematic literature review for blockchain.

| Paper | Cutoff Time | Keyword | Sources | Unscreened papers | Papers Selected | % of Non-Bitcoin | Questions | Comments |
|---|---|---|---|---|---|---|---|---|
| [51] | 2015 (estimated) | *cryptocurrency, cryptocurrencies, crypto AND currency, crypto AND currencies and Bitcoin* | ABI/Inform, ACM, AIS eLibrary, IEEE, Certain IS conferences and IS journals | 54 | 42 | Mostly Bitcoin | 1. Which methods, concepts, ideas and approaches of cryptocurrencies have been researched in scientific literature? 2. Which IS research areas can be linked to cryptocurrency research and what are future research topics for cryptocurrency related IS research? | Focus on Bitcoin and crypto currency |
| [47] | 2016/5 | *blockchain* | IEEE Xplore, ACM Digital Library, Springer, ScienceDirect, Ebsco, PLOS One | 121 | 41 | 80% Bitcoin | 1. What research topics have been addressed in current research on Blockchain? 2. What applications have been developed with and for Blockchain technology? 3. What are the current research gaps in Blockchain research? 4. What are the future research directions for Blockchain? | Focus primarily on Bitcoin and its technical aspects especially security and privacy |
| [48] | 2016/12 | *blockchain* | Springer, PRL and Google Scholar | Not Available | 54 English, 216 Chinese | 83% on Bitcoin and cryptocurrency | What are the research subjects (financial, credit, accounting, and others) and research methods (qualitative vs. quantitative) and future directions blockchain related papers? | Focus on Bitcoin and cryptocurrency |
| [53] | 2017 | *blockchain, block chain, peer-to-peer database, immutable database, consensus database, consensus protocol, distributed ledger* | Google Scholar | Only the first 50 search results of each search are analyzed | 31 | Not available | What are the characteristics of blockchain technology and its impact on service systems? | Focus on identifying blockchain characteristics on trust and decentralization, and their impact on service systems. |
| [31] | 2017/01 | *blockchain, block chain* | Web of Science, IEEE Xplore, AIS Electronic Library, ScienceDirect, and SSRN | 3792 | 69 | Mostly non-Bitcoin | What is the current state of knowledge regarding blockchain, and how can it purposefully be advanced? | Focus on a general blockchain literature review; propose a blockchain framework adapted from well recognized social media research agenda to stimulate multidisciplinary research approaches |
| [49] | 2017 | *blockchain* | IEEE Xplore, Springer Link, ScienceDirect, the YMCA, Google Scholar | Not available | 190 (including both English and Portuguese papers) | 60% non-Bitcoin blockchain and 40% Bitcoin | 1. What is the evolution over the years in the number of publications on blockchain? 2. What are the main features of research analyzed in research on blockchain? 3. What are the application areas of blockchain technology? 4. What are the limitations in current research in blockchain research? 5. What are the future trends and challenges to search for blockchain? | Focus on reporting the number of documents in four categories: IoT, Smart Contracts, E-governance, and Others, with brief discussion on each. |
| This paper | 2018/6 | *blockchain* | ACM, ASCE, IEEEXplore, EGRL, JSTOR, ScienceDirect, Scopus, Springer, Web of Science, Wiley. | 3827 | 159 & 71 | Mostly non-Bitcoin | 1. What are the blockchain use cases studied by the research community in the sectors central to smart and sustainable cities? 2. What could be a unified framework to examine those use cases regardless of their sector? 3. How do we evaluate the impact of the use cases on urban sustainability? 4. How do we evaluate the blockchain applicability of the found use cases? 5. What appropriate taxonomies of blockchain applications can be derived to facilitate cross-sector use case analysis? | Focus on an application oriented use case review under a sustainable and smart city context; propose component-based analysis framework and business model-based taxonomies for cross-sector use case analysis |

**TABLE 3.** Paper list for application-oriented use case review.

| Sector | List of Papers | No. of Papers |
|---|---|---|
| Governance and citizen engagement | [5], [61]–[78] | 19 |
| Education, culture, science and innovation | [79]–[102] | 24 |
| Well-being, health and safety | [103]–[130] | 28 |
| Economy | [131]–[150] | 20 |
| Transportation | [151]–[172] | 22 |
| Energy | [173]–[206] | 34 |
| Water and waste management | [207] | 1 |
| Built environment | [208]–[214] | 7 |
| Natural environment | [215]–[218] | 4 |
| Total number of papers | | 159 |

It is important to note that active engagement of citizens and stakeholders in collaborative urban governance is hardly political in nature [223]. That is because collaborative urban governance taps into the intelligence of all urban actors to create public values by providing conditions to motivate knowledge generation, exchange and innovation by citizens [229]. This way, the citizens themselves become the best regulators of cities [219]. Open data offers an example of strengthening collective intelligence of city stakeholders to derive innovations, even though at the same time governments should carefully decide how and to which actors this data is opened up [230] and how to protect the confidentiality, privacy and intellectual property rights for data and model development [225].

The blockchain use cases in the surveyed papers tend to discuss solutions on the more progressive side of the digital governance spectrum. In particular, using innovative IT to transform existing processes and better serve citizens (the "smart administration" model) and collaborative urban governance (the "smart urban collaboration" model).

### 1) INNOVATIVE IT TRANSFORMATION FOR EXISTING PROCESSES

Under this category, blockchain-based systems have been proposed to transform the government document sharing process. Between the government and the public, [61] describes a system that stamps a government decision on the blockchain to keep an immutable and transparent record to be verified anytime in the future; between government and businesses, [62] designs a system for business to share information with government organizations in a way that helps business both ensure the confidentiality of the information and avoid liability; between government agencies themselves, [63] presents an inter-agency document sharing system, where the requesting agency first looks up a pre-constructed catalog to locate the destination agency and then conducts a direct document sharing exchange with it. The transaction is at the same time recorded on the blockchain for robust and secure access control.

There are also blockchain systems targeting at transforming two of the most important government processes - one is voting, which forms the government, and the other is taxation, which finances the government. The goal of an E-voting system is to achieve anonymity, privacy and transparency [64]. Anonymity ensures the non-traceability of the voter's vote. Privacy allows the transaction of the vote to stay hidden, and transparency ensures the public that the voting mechanism cannot not be tampered with. Design of blockchain systems for voting are found in a number of literatures [64]–[69]. Some also produced prototypes [70]–[72]. However, an issue with all these systems is that the authentication of voters at the personal level has to be ensured outside the blockchain.

In the area of taxation, blockchain solutions enable tax authorities to have more control over the tax system. Reference [73] describes a private blockchain that can be managed by the tax authority to monitor value-added-tax invoices and keep an immutable record on the taxable transactions, thus preventing tax revenue losses. The system in [74] tackles a different scenario by using the blockchain to track the dividend paid to stakeholders, in order to overcome the problem of duplicated tax refund due to forged dividend payment claims.

### 2) CITIZEN-CENTRIC COLLABORATIVE URBAN GOVERNANCE

While the urban collaboration perspective is the dominance of transformational ideas in literature on smart city governance [223], an inherent issue is that the e-governance models associated with smart city initiatives typically rely on Internet-based online tools which are increasingly monopolised by a few companies serving as de facto central authorities [231]. Lack of transparency and trust on a centralized network infrastructure could be a key factor that hinders the true realization of the citizen participatory governance model.

Researchers believe that blockchain has the ability to decentralize the Internet [231], enable a decentralized delivery model that allows rethinking complex systems in a more participatory manner [232], and become an important infrastructure for e-government [233]. Blockchain helps build societal trust with an intrinsic checks and balance systems and promotes a society of dignity, recognition, and respect [234] that could be fundamental to the most transformative collaborative governance model.

One project illustrating this vision is [5] which concentrates on the area of urban policy making. The authors state that current urban codes such as policies, planning, regulations and standards are not up to meeting the urban sustainability challenges due to their top-down delivery and implementation methods. Blockchain-based mechanism makes it possible to truly deliver and execute urban codes bottom-up. A blockchain system [5], [75]–[77] was proposed as a connecting mechanism to create the people's layer of the governance systems that connects urban technologies. In the case of policies and codes, citizens submit their urban needs to the blockchain, which will be prioritized by a blockchain consensus mechanism for the authorities to draft policies. These drafts will be ratified through the blockchain validation capabilities. Further transform of these plans into physical forms (e.g., construction of infrastructure projects) can be approved via voting mechanisms on the blockchain as well. The plans and regulations can also be standardized for replicability and scalability purpose, using the same bottom-up approach for citizen participatory standardization.

Another related blockchain project shows a citizen-participatory decision support framework under the healthcare context [78]. It runs an agent-based simulation model on the blockchain, incorporating rules from expert stakeholders, open data and anonymized volunteer participants data. The project illustrates how such a framework can be helpful in an infectious disease spread scenario, through improving transparent and ethical management of individual data and promoting evidence-based collective decision making.

### B. EDUCATION, CULTURE, SCIENCE AND INNOVATION

#### 1) EDUCATION AND LEARNING ACTIVITIES

Blockchain systems have been proposed to help maintain an immutable record of the educational process. There are proposals that record creative works or ideas to establish scholarly reputation [79], create continuous log of the learner's activities across different learning organizations [80], enable global higher educational institutions to award course credits to students who completed courses [81], and allow issuance and revocation of educational certificates [82]. Education and other records can be inputs to a general personal archive management system and used by companies and services for verification [85].

A related topic is recording of broader learning activities such as volunteer services. Reference [83] describes a blockchain based system for life-long volunteering. Unlike other systems that focus on scheduling and allocation of tasks, it fosters an open volunteering marketplace supporting intelligent matchmaking, gamification, and goal-oriented personal development. Blockchain serves to store the persistent digital footprints for volunteering activities, assessments and acquired qualifications, and also gives data sovereignty to the volunteers themselves. Reference [84] is another system that discusses using blockchain to record volunteer service time and activity information.

#### 2) SCIENCE, INNOVATION AND IP PROTECTION

Scientific researchers are also using blockchains to solve problems associated with the academic community. The use cases cover the whole life cycle from research methodology, peer review, manuscript publishing to intellectual property protection. First, at the experimentation stage, to prevent experimental integrity from being damaged by negligence or intentional wrong-doings, [86] proposes a system to record research datasets and results to blockchain and release them when necessary, e.g., upon approval of multiple specified signatories, thus providing an audit trail of research data. Reference [87] suggests using adaptable blockchain-based choreographies for collaborative, reproducible in silico experiments towards both Robust Accountable Reproducible Explained (RARE) research [235] and Findable Accessible Interoperable Reusable (FAIR) results. Second, at paper authoring stage, [88] presents a blockchain platform that preserves and measures author contributions based on the edits that authors commit. Third, at the peer-review stage, blockchain system can also stimulate a timely and sustainable review process. As described in [89], a system can reward cryptocurrency to reviewers when a quality review is accepted by the editor. These rewarded currencies could later be used to pay for publishing the reviewer's own authored paper in journals, forming a closed loop incentive mechanism. Forth, at the publishing stage, [90] leverages prior work of semantic web technologies to allow authors to collaborate on an evolutionary version of the research progress, which could be open for reviews or submitting to conferences or journals. This provides the possibility for a decentralized publishing system (in contrast to the existing system centered on major publishers). By ensuring a single version of truth throughout the paper life cycle, the system can solve the trust issues among the different actors in the publishing ecosystem, including authors, reviewers, publishers and relevant personnels who use bibliometrics to evaluate performance. Last but not least, on intellectual properties, [91] reports a blockchain system that automatically creates a publicly verifiable timestamp for each submitted manuscript, facilitating its origin time record protection. Use of blockchain for intellectual property protection extends to software as well. References [92] and [93] present design of a software licensing validation system for publishers, enterprises and end users. Using ownership of crypto-tokens to represent software entitlement, the blockchain enables license validation, software updates, license transfers and related functions. Reference [94] discusses a system for licensing of 3D printing models. It links the model to license data on the blockchain in order to secure the authenticity of printing data and prevent its unauthorized use.

#### 3) MEDIA, CULTURE AND ENTERTAINMENT

Intellectual property protection for digital media is also a common blockchain application [95]. The system in [96] proposes to register self-embedding watermarking processed images on the blockchain in order to preserve transaction trails and content modification histories, and provide tamper detection for digital image management and distribution. Reference [97] reports a system for multi-media rights management, allowing the licensor to control permission of particular licensee to play the designated videos. Blockchain technology application transforms the roles of third party intermediaries in the media industry, making artists' careers more sustainable by improved overall transparency of the value chain [98]. But there are also cautions on the feasibility of blockchain in the copyright sphere [99]. Reference [100] argues that using blockchain as a financialisation tool through media rights enforcement is unlikely to empower artists but instead will curtail the critical potential of the digital as a mode of production and artistic expression.

Examples for blockchain in cultural and entertainment area include [101] which implements a blockchain system that can securely manage transfer, re-selling, validation of concert event tickets to prevent ticket fraud, and [102], a decentralized lottery system to ensure fairness, transparency and privacy of the lottery process.

### C. WELL-BEING, HEALTH AND SAFETY

Healthcare is a prominent area where blockchain could found many use cases that help establish an infrastructure to ensure transparency of medical data, analytical methods, reproducibility of results and improved trust in translational medical value chain [60]. As such, they have the potential of significantly reducing the cost of developing new drugs, diagnostic tools, and clinical regimes [236]. The following main categories of use cases are found in the research papers.

#### 1) CLINICAL TRIALS AND MEDICAL RECORDS

Blockchain systems have been proposed to improve the clinical trials process, from keeping track of each steps, such as patient consent and any revision of the clinical trial protocol [103], [104], to managing complex clinical trial data and preventing them from unauthorized manipulation [105].

Many papers presented concepts and designs of systems that target at enabling secure, interoperable, and efficient access to medical records by patients, providers, and third parties. They place emphasis on access control challenges associated with sensitive data storage [106], preserving data privacy and integrity [107], storage and retrieval [108] and cross-domain medical image data sharing [109]. Blockchain prototypes of such systems have been reported both for general purpose medical records [110], [111] and for particular medical areas such as oncology patient care [112]. In addition, [113] presents a system specifically for medical data access control between cloud service providers. The effort in [114] ensures data access accuracy for public reference biomedical databases by providing query notary.

Some proposed systems also include collecting medical data in mobile environment. The systems in [115] and [116] generate health record data from patient's smart devices and register them to the blockchain network for tamper-resiliency. The concept of applying blockchain in pervasive social network based healthcare is explored in [117].

#### 2) DRUG AND FOOD SUPPLY CHAIN

Medical supply chain may benefit from blockchain technology to help protect public health. Such systems can trace the origins of drugs by logging time series drug transaction data generated by IoT sensors to a blockchain to prevent counterfeits [118]. The recording of drug transactions among manufacturers, wholesalers, retailers, pharmacies, hospitals, and consumers, can turn the drug supply chain from regulating (government audits) to surveillance (by every participants collaboratively) [119]. Another system [120] focuses more specifically on real-time tracking of all cannabis plants from their production to final destination in order to undermine their illegal markets.

There are also systems that tackle the transport aspect of medical products. Medical products require specific quality control and regulatory compliance such as assertion of temperature and humidity during the transport process. [121] built a system using IoT sensors to collect those transport condition parameters and log the readings to a blockchain for public verifiability. Blockchain is also a hot topic for tracing of food and agricultural products [122], [123]. Researchers have focused on secure data storage scheme for blockchain-based agricultural product tracking systems [124], presented customized blockchain for agricultural resource supply chain [125], and discussed case study of blockchain agriculture and food traceability in China [126].

#### 3) INSURANCE

Improving the insurance sector is also what people believe blockchain can be helpful. Reference [127] discusses applying blockchain to the insurance life cycle, from seeking a quotation to binding a policy contract, to the claiming process, which could help reduce fraudulent insurance claims. Experimental prototypes have been created to offer fine-grained insurance policy control [128].

There are specific subjects of insurance that received more attention. Reference [129] implemented a micro-insurance use case for managing and analyzing data in a pay-as-you-go car insurance, which allows drivers who rarely use cars to only pay insurance premium for particular trips they would like to travel. Reference [130] built a blockchain-based prototype for cyber insurance. The system aims to provide an automated, real-time, and immutable feedback loop among the involved parties, providing a secure distributed infrastructure for assessing cyber risks.

### D. ECONOMY

While we do not consider pure cryptocurrency use cases in this paper, nor do we discuss Initial Coin Offerings [237], we list the following important areas we found in the surveyed literature that blockchain is affecting the broader economy domain.

#### 1) COLLABORATIVE BUSINESS PROCESSES AND SERVICE EXCHANGES

Blockchain has great potential in business process management toward building a distributed, trustworthy infrastructure to promote inter-organizational processes [238]. Reference [131] describes a decentralized social manufacturing platform where prosumers publish service demands and the manufacturing community works to satisfy the demands. Similarly, [132] presents a case on collaborative fulfillment of industrial product design. Another system is discussed in [133] which provides a platform for software development and automatic payments, incorporating trusted Oracles for automatic software code verification.

Blockchain facilitated collaboration can be performed not only by human, but also by autonomous agents. Reference [134] illustrates such a scenario of organizing a network of unmanned aerial vehicles to make scheduled delivery flights and report on the mission performance.

### 2) E-COMMERCE

Regarding the commerce market, a peer-to-peer blockchain-based e-commerce platform is reported in [135] and used by employees of a large multi-national company. Blockchain efforts are also used to fight counterfeit goods in commerce. Reference [136] describes the design of recording ownership of products on blockchain. Reference [137] implemented a blockchain-based prototype for product ownership management for the post supply chain so that counterfeiters can be detected by the consumer.

The data integrity feature of blockchain has been used for other purposes in online commerce, specifically, [138] uses it to prevent forged digital ads clicks for fraudulent service commissions. The system enables user to link several ad reports together in a form that resembles the architecture of a blockchain. This blockchain, together with incorporated user social behavior patterns, allows advertisers to identify authentic ad reports.

Aside from online commerce, automated physical sales systems such as vending machines could also leverage blockchain technology. In the system described in [139], automated sales systems record product quantities and sales information on the blockchain, so users can always obtain the current product information of the systems.

Another direction in the commerce market involves machine-to-machine payments and human-machine hybrid payments. The machine-to-machine payment system presented in [140] enables a smart cable connected with a smart socket to pay for electricity using Bitcoins without any human interaction. In order to alleviate the high Bitcoin transaction fee problem, it uses a single-fee micro-payment protocol that aggregates multiple smaller payments incrementally into one larger transaction. A hybrid transaction interaction between human and machine is reported in [141]. It proposes a conceptual design of a blockchain system to ensure the integrity and non-repudiation property of messages controlling a smart door lock. The smart door lock verifies received control messages for authenticity and records any door control transactions on the blockchain. Reference [142] is another work involving hybrid financial transactions between a robot and a human for the robot to complete assigned tasks and have the outcome asserted on the blockchain.

### 3) REPUTATION SYSTEMS

Reputation mechanism is important in a commerce market and they can also benefit from blockchain technologies.

References [143] and [144] discussed the design of a blockchain-based binary reputation system for file transfer transactions, where the rating could be either 1 for positive or 0 for negative. A more general blockchain-based reputation system for e-commerce applications is presented in [145] which allows customers to leave text reviews. The service providers need to earn and spend crypto tokens in order to receive a review from a customer. The reviews are recorded in the blockchain to ensure temper-resiliency while eliminating third party intermediaries. Reference [146] is another blockchain-based reputation system and it aims at using a single protocol to achieve efficient, anonymity-preserving, decentralized, and robustness against various known attacks such as ballot-stuffing and Sybil attacks.

### 4) SHARING ECONOMY

Sharing economy can also be boosted by blockchain's capability to promote trust-free transactions. Reference [147] is a system for sharing any kind of everyday tangible object. It enables peer users to rent devices (e.g., power tools) without disclosure of any personal information. Another peer-to-peer market prototype is reported for leftover foreign currency exchange [148] that could help alleviate the challenges of bringing leftover foreign currency back into circulation. Intangible resource sharing has also been studied. Reference [149] presents the design of a blockchain system for citizen broadband radio service spectrum sharing. The system could significantly reduce operational costs, introduce flexibility and scalability into spectrum regulation, and allow new entrants to access local spectrum based on their specific business needs.

To unleash the full potential of the sharing economy, some people experimented a more social relations-based production model, as exemplified by the Backfeed project [150]. It develops governance and economic models for decentralized organizations to enable collaborative economy using blockchain. In this framework, people contribute to a common effort, evaluate each other's contribution and achieve decentralized consensus on the produced value. Fair share of the created value and rewards for the contributors are presented through a crypto token based economy. The blockchain maintains a permanent, transparent, and secure infrastructure for the overall ecosystem.

### E. TRANSPORTATION
#### 1) VEHICLE INFORMATION MANAGEMENT

Vehicle life cycle information is critical for the huge car markets. Reference [151] presents a blockchain-based system for recording and managing vehicle data to increase the transparency, reduce odometer and other frauds.

#### 2) GOODS TRANSPORTATION

In the goods transportation area, there are many discussions on blockchain as a way to digitize the exchange of shipping documentation, bill of lading and compliance [152], [153], all holds potential to cut costs in global trade. Reference [154] presents a customized blockchain implementation that supports tamper-proof traceability of data and automates regulatory compliance checking. Reference [155] offers another blockchain-based prototype for cargo tracing capability and various supply chain management tasks. Some other work emphasizes tracing goods of specific types and sectors, such as dangerous goods [156], aircraft parts [157], or the marine sector [158]. There are also efforts that combine blockchain with different identification technologies such as Near Field Communications (NFC) tags [159] or with Krakelee fingerprint [160].

### 3) INTELLIGENT TRANSPORTATION SYSTEMS

Intelligent transportation systems use information and communication technologies to improve efficiency in road transport, traffic management and mobility management, as well as for interfaces with other modes of transport [239]. While most intelligent transportation systems are centralized, blockchain has been proposed to help create a secured, trusted and decentralized autonomous intelligent transportation ecosystem, allowing the control and management of both physical and digital assets [161].

Reference [162] presents the concept of distributed transport management system for vehicles to share their resources and create a network in which they can produce value-added services, such as automatic gas refill and ride-sharing. Online taxi-hailing and ride-sharing is considered a prominent application scenario in this area. Unlike other solutions like Uber and Lyft, a blockchain-based solution allows better personal privacy protection as the taxi software platform cannot obtain the entire itinerary of the user; the user can also control access to his travel data records [163].

Vehicle-to-Everything communications is a key component in the proper functioning of an intelligent transportation system. Many efforts have been conducted on security of these communications. Security credential management systems [240] are created to issue certificates to trusted vehicles and revoke certificates of misbehaving ones to ensure message security and privacy. Reference [164] describes the design of a decentralized alternative to existing security credential management systems by using blockchain technology to remove the need for centralized trusting authority. It improves the global revocation algorithm performance through hierarchical consensus, and creates accountability for misbehaving parties. The system in [165] tackles dynamic key management for heterogeneous intelligent transportation systems by leveraging the blockchain network to transport vehicle security keys across different security domains. References [166] and [167] proposed blockchain-based reputation system in vehicular networks. Vehicles rate the received messages based on observations of traffic environments and store them on the blockchain. These ratings represent the consensus of crowds on each vehicle's reputation and allow vehicles to assess the credibilities of received messages.

Other efforts, [168] and [169] also focus on secure inter-vehicles communication mechanisms.

Reliability of the communications, such as privacy-preserving incentive announcement network based on blockchain is examined by [170]. Through an efficient anonymous vehicular announcement aggregation protocol, the system helps improve the reliability of announcements in the non-fully-trusted vehicular ad hoc network, without revealing users' privacy.

Software updates for smart vehicles is also important to keep the system up-to-date and secure. Reference [171] proposes blockchain-based security architecture to perform over-the-air updates for smart vehicles remotely, or to securely distribute the latest software to service centers for them to be installed on a vehicle locally.

### 4) URBAN TRANSPORTATION SUSTAINABILITY

Reference [172] implemented a blockchain-based financial incentive system to encourage urban cycling. It allows cyclists to collect their activity data and monetize their commuting habits through the blockchain, thus encouraging sustainable transport in cities.

### F. ENERGY

In the energy domain, electricity and the grid have been a focus for blockchain related applications.

### 1) GRID SECURITY AND METER TRANSPARENCY

Reference [173] proposes a framework that harnesses blockchain's distributed features to enhance data security in a network of smart meters. Signed meter reading messages are broadcast to and validated by peers, and then recorded to a private blockchain. Research also used blockchain to provide a consumer-facing utility usage monitoring system in order to help customers understand how the appliance are consuming electricities and be sure that utilization data cannot be artificially manipulated [174], [175].

### 2) PEER-TO-PEER ENERGY TRADING

Mass deployment of rooftop solar photovoltaic cells is shifting electricity consumers to producers-consumers (prosumers), much like citizen journalists in social media [241]. These prosumers seek to both reduce their power bills and to sell their excess power to others, creating a new business model on peer-to-peer energy trading transactions [176]. Such transactions improve resiliency in the grid and offer the possibility of exchanging distributed energy at speed, scale and security [177].

There are extensive research discussions on blockchain enabling peer-to-peer energy trading with the smart grid and micro grid [242]. Some research efforts focus on demand and generation balance in the grid network. Reference [178] proposes a decentralized optimal power flow model for scheduling a mix of batteries, shapable loads (e.g., electric vehicles with continuous charging levels), and deferrable loads (e.g., appliance and manufacturing equipments) on an electricity distribution network. The goal of that blockchain-based system is to maximize social welfare by scheduling the controllable loads to minimize generation cost while respecting network constraints. Reference [179] presents a blockchain-based energy system for automated negotiation, settlement and payment, plus reward for system demand supply balancing support. Prosumers submit their demand request and supply offers. The system determines whether those offers are accepted (e.g., if demand is higher than supply then the supply offers are accepted in order of lowest to highest). The actual usage and supply of the users are measured. Those who help the system resolve imbalance are rewarded and those who make the system more imbalanced are penalized. Reference [180] is a work focusing on efficient use of shared energy resources to minimize external dependence. It provides a blockchain-based distributed controller. Through its coordinated operation, the energy storage systems of households in local energy communities can achieve an increase in efficiency and self-sufficiency. The system in [181] emphasizes the regulation of energy production and distribution, and give specific attention on discouraging the production of non-renewable energy. Reference [182] is a system that centers on demand side management of the grid. The prototype in [183] uses actual energy traces of UK building datasets to validate a blockchain based decentralized management system.

Some other work examined different market mechanism and agent behaviors in the peer-to-peer electricity trading system. Reference [184] studies the double auction market, which collects bids over a specified time interval and clears the market at the end of bidding interval. It implements a zero-intelligence agent bidding strategies where the agents randomly quote within a uniform distribution without considering market transactions. Since micro grid transaction cycle could be short due to uncertainty of renewable energy power generation, the work in [185] evaluates a continuous double auction, which matches buyers and sellers immediately upon the detection of compatible bids. It also adopts adaptive aggressiveness for agents, enabling them to adjust quote automatically through learning mechanism according to market price and price fluctuations. In addition, [191] implements a blockchain-based platform extending the features of cryptocurrency exchanges to provide a robo-advisor like system to recommend the best selling strategy for prosumers in the renewable energy market.

The overall peer-to-peer electricity trading framework has been explored a lot. Reference [186] proposes a system that targets at various typical industrial IoT scenarios, such as micro grids, energy harvesting networks, and vehicle-to-grids. A credit-based payment mechanism is included to support fast and frequent energy trading, alleviating the low throughput problem in typical blockchains. The trading system presented by [187] connects energy producer smart meters and local battery/AC main as a distributed energy network. A controller middleware bridges communications between the physical energy network and the blockchain. Reference [189] discusses a machine-to-machine electricity market in the specific context of the chemical industry. References [190] and [192] also present peer-to-peer electricity trading architectures. Another work, [188] concentrates on implementation details of a blockchain electricity trading system. Research also shows that homeowners want to preserve their privacy in using local sources of energy [193]. Therefore, privacy and anonymity is also a focus area of many papers [194]–[197].

While most of the systems described in literature are designs or research prototypes, the Brooklyn Microgrid, a micro grid energy market in New York is an actual test bed in the operation. In that system [198], the micro grid serves as a backup that can be decoupled from the traditional grid in case of power outage. The users' electricity consumption and generation data is logged in their blockchain accounts and electricity transactions are conducted through blockchain-based market mechanisms.

### 3) ELECTRIC VEHICLE AND GRID
Using blockchain for electric vehicle and grid is another intensive area of investigation. Reference [199] discusses a system targeting at guaranteeing the execution of energy recharges for the autonomous electric vehicles refueling scenario that meets the requirements of latency, security and cost. Reference [200] is a blockchain-based design to enable electric vehicles to autonomously select the most appropriate charging stations among list of bids according to, e.g. the planned route, car battery status, real-time traffic information and drivers' preferences. The protocol in [201] allows electric vehicles to find the cheapest charging station within a previously defined region and preserve the privacy of the electric vehicle. Prototype systems for electric vehicle and grid charging are presented in [202] and [203].

In addition, [204] discussed minimizing the power fluctuation in the grid and reducing the overall charging cost for electric vehicle users; [205] examined charging scenarios involving mobile charger for electric vehicles. Peer-to-peer electricity trading system directly between plug-in hybrid electric vehicles is studied in [206].

### G. BUILT ENVIRONMENT
The built environment and the Architecture, Engineering and Construction (AEC) industry have also been covered in blockchain application discussions. Trust, information sharing, and process automation are of great value in construction engineering [208]. Trust relations in the construction industry concern about people from organizations such as clients, contractors, subcontractors, and suppliers [209]. An "Evidence of Trust" is especially important in the AEC industry because it can scaffold the collaboration between the involved parties, and true collaboration is critical for design and construction [210]. The trust and many other core issues in this industry are rooted from the distributed and complex nature of construction projects. Solving these issues would unlock capability and productivity of the AEC industry - just as innovative socio-technical mechanisms of past centuries led to explosive advances in global trade and communication [211].

Nevertheless, blockchain applications in the built environment and AEC fields are relatively less explored by the research community. Even for the few papers we were able to find, most of them are in the conceptual discussion and design phase. Reference [208] proposes possible scenarios for blockchain use in the construction industry which includes notarization applications to eliminate the verification time of construction documents authenticity, transaction applications to facilitate automated procurement and payment, and provenance applications to improve transparency and traceability of construction supply chains. Reference [212] discusses more specifics on multi-party automated and performance-based payment upon construction completion. The process could be integrated with Building Information Modeling (BIM) and sensor-based remote monitoring as well as visual data analytics. More on blockchain integration with BIM is discussed in [213]. Taking a different perspective, [214] explored using blockchain to enhance access control in building operating systems that are designed for energy efficiency, human comfort, and grid integration of buildings. Leveraging a blockchain-based authorization syndication platform, the authors built a prototype that extends the building operating system beyond the single administrative domain of a building, to enable democratized delegation of authorization in multiple administrative domains without centralized trust authority.

### H. WATER AND WASTE MANAGEMENT

Related to water consumption efficiency, [207] proposes a privacy-friendly blockchain-based gaming platform that aims at engaging users in diminishing water or energy consumption at their premises. Teams can compete with each other or against unmanned adversary. Through collection of secure commitments from the utility meters, the blockchain mechanism allows the users to formally prove that they have correctly reported their measurements without disclosing the measurements themselves in order to preserve privacy.

### I. NATURAL ENVIRONMENT

Use cases found in natural environment include air quality monitoring, sand resources management and carbon credit trading. Reference [215] proposes a blockchain-based system that encourages the constructive involvement of urban citizens in monitoring environment quality to promote a greater awareness of city health. Reference [216] suggests that sand should be treated as a key resource on a par with clean air, biodiversity, and other natural endowments. It describes a blockchain-based approach to monitor the supply chain of sand resources from mining to trading in order to prevent illegal sand mining.

Blockchain technology and smart devices can also be used to improve carbon emission compliance and trading. Reference [217] presents design of a blockchain-based emission trading system for the fashion apparel manufacturing industry. The system aims at reducing the emissions for all the key steps of clothing making and involves the authority, the auditors, the firms as well as the related individuals. Another blockchain-based emission trading proposal [218] incorporates a reputation-based mechanism to encourage the participants to adopt a long-term solution in emission reduction.

## VI. COMPONENT-BASED BLOCKCHAIN USE CASE ANALYSIS

In order to provide an anatomy of typical blockchain use cases, we define a general analysis framework as shown in Figure 1. At the top part are the writers and readers who interact with the blockchain to update or view records. At the center are the assets, which are the subjects of the blockchain records. Depending on the use cases, three key properties about the records have to be considered: transparency, privacy and anonymity. At the bottom part of the framework stand three important pillars for the underlying blockchain platform: distributed consensus mechanism for the database, smart contracts for business logic, and crypto tokens used by the infrastructure or applications. In terms of the relationship among these components, the writers and readers on top are external to the blockchain and the bottom pillars are internal to the blockchain platform. The assets in the center are the linkage between these two parts, as there could be both off-chain and on-chain assets with appropriate mappings. The transparency, privacy and anonymity requirements of the assets come from the external use case properties but are fulfilled by the internal blockchain infrastructure.

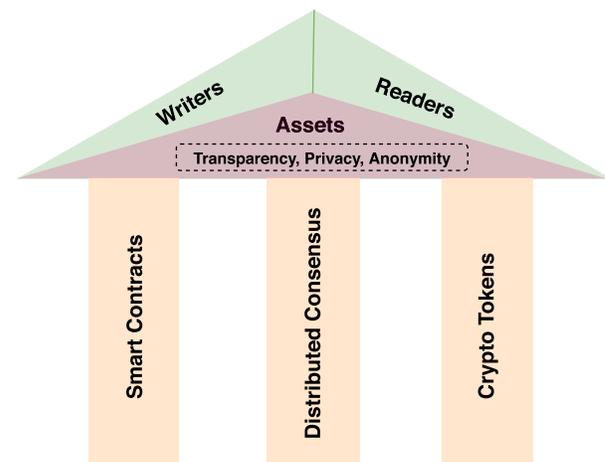

**FIGURE 1.** Component-based blockchain analysis framework.

A practical difficulty we face is that not all surveyed papers explicitly discuss all aspects of the above framework in their work. This is common because many of them may be focusing on other specific aspects of the use case, or are at a design phase where not all system decisions need to be made.

Therefore, we have to select papers that provide enough information regarding the components of the framework. The chosen subset of 71 papers is listed in Table 4. They represent

| Sector | List of Papers | No. of Papers |
|---|---|---|
| Governance and citizen engagement | [5], [61], [63], [70], [71], [73], [74], [78] | 8 |
| Education, culture, science and innovation | [81], [82], [88], [90], [91], [93], [96], [101], [102], [243] | 10 |
| Well-being, health and safety | [105], [107], [109], [110], [113]–[118], [121], [124], [128]–[130] | 15 |
| Economy | [132]–[135], [137]–[140], [146]–[148], [150] | 12 |
| Transportation | [151], [155], [165], [172] | 4 |
| Energy | [173], [175], [178]–[180], [183]–[189], [195], [197]–[199], [202], [204], [206] | 19 |
| Built Environment | [214] | 1 |
| Natural Environment | [216], [217] | 2 |
| Total number of papers | | 71 |

all the prior sectors except the "Water and Waste Management" sector. Most of them present system prototype implementations. Some of them do not have explicit implementations but still provide fair amount of details on the system that allow us to deduce the components they use. We include them to ensure a reasonable dataset size.

## A. ASSETS

In the core of all blockchain applications is a distributed database that keeps a tamper-resistant record for the associated assets. A key characteristic of the use cases we surveyed is that they all involve off-chain assets. This is different from pure cryptocurrency blockchain applications, which may involve on-chain cryptocurrency assets only.

We produce a consolidated list of the off-chain assets appeared in our studied set of use cases in Table 5. In this table, we categorize the assets into digital or non-digital. We consider digital assets to include both assets that are digital by origin such as software, digital media, and those that could have a native digital representation, e.g., an electronic version of a document or a lottery ticket. Non-digital assets may be tangible (physical) or intangible, both need to be digitalized before they can be brought on-chain.

For physical assets that are distinguishable from each other, their commonly known unique digital identifier can serve as their natural representation on the blockchain, such as the EPC of goods [137], VIN of vehicles [151], RFID tag of cargo [155], or ORCID of research authors [90]. Usually, digital representation of the assets include additional attributes depending on the use case, such as the ownership of products [137], the mileage number of vehicles [151], the location and temperature conditions of the goods in transportation [121], [155].

Non-distinguishable physical assets such as natural resources (e.g., water and sand) can be digitalized with their ownership and value attributes. For example, meter reading for water consumptions [207] and sand mining demand amount [216]. These values are also associated with the identity that is responsible for those resources like the water meter owner and the sand miner.

Intangible assets can also be digitalized through ownership and values. Electricity is a key non-digital intangible asset found in the energy sector. It is usually represented by digital readings from smart meter devices (e.g., [173], [175], [186], [187]), and the identity of the meter device links the electricity asset to its owner.

Once assets become digital (either native or converted), they still need to be brought on-chain through human or machine based operators. However, there could be an additional step before the asset is taken on-chain. Blockchains are known to be notoriously unsuitable for storing large files directly. This is because the mechanism that ensures the blockchain's immutable and tamper-resistance properties necessitates a lot of expensive cryptographic computations for on-chain transactions. Even for very small transaction data size like that in the Bitcoin blockchain, the resulting throughput is much slower than that of comparable non-blockchain platforms. Therefore, only certain small sized digital assets may be recorded onto the blockchain directly, for example an event ticket [101], lottery ticket [102], number of completed course credits [81] or a software model license code [94]. For majority of the assets that require larger space, e.g., medical images [113], government policy document [61], the best practice is to store a cryptographic hash of the original asset as its verified proof. The decision of whether to store the asset in full or in hash format is dependent on many practical factors such as what kind of blockchain platform is used and what performance result is sought. For the sake of brevity, we will not explicitly state whether the asset is stored in full or in hash format in our discussion of use cases for the rest of this paper.

## B. WRITERS

In our analysis framework, the writers refer to parties who can submit changes to the blockchain database. It is important to note that the writers are external actors to the blockchain use case. They can be separate from the actual blockchain nodes that validate the transactions, reach consensus and commit the transactions records onto the blockchain. In other words, writers submit transactions that could update the blockchain database, but it is up to the blockchain's internal mechanisms to accept those updates.

The records that writers submit to the blockchain describe asset attributes. They can be static attributes, e.g., the authorship of a research manuscript [88], or dynamic values resulting from continuous monitoring, e.g., DarkWeb status reporting for Cyber security insurance [130]. An important common category of records involve asset exchange transactions, e.g., in commerce market [135], [137]. These records are essentially a special case of dynamics asset ownership attributes.

Permission to write and symmetry of writing privileges among all writers are important factors that differentiate the use cases. A use case could allow either public writing

 **Consolidated list of assets from the surveyed papers.**

| Sector | Assets | Asset Type | Sample Records |
|---|---|---|---|
| Governance and citizen engagement | Government policy document | digital | Government policy documents |
| | Vote | digital | Vote |
| | Tax invoices | digital | Tax credits |
| | Disease spread information | digital | Disease spread report |
| Education, Culture, Science and Innovation | Research paper and review | digital | Research paper and review |
| | Software | digital | Software license |
| | Educational certificates | digital | Educational certificates |
| | Course credits | digital | Course credits |
| | Digital media | digital | Digital media license |
| | Event ticket | digital | Event ticket |
| | Lottery ticket | digital | Lottery ticket |
| Well-being, health and safety | Healthcare record data | digital | Healthcare record data |
| | Drug | physical | Drug transaction |
| | Food | physical | Food transaction |
| | Insurance | digital | Insurance monitoring information |
| Economy | Digital products | digital | Service outsourcing transaction |
| | Physical goods | physical | Goods ownership and transaction |
| | Transaction review | digital | E-commerce review |
| Transportation | Vehicle | physical | Vehicle information and transaction |
| | Physical goods | physical | Goods transportation |
| Energy | Electricity | non-digital intangible | P2P energy demand/supply and transaction |
| Built Environment | Building resources | digital | Access control delegation for building resources |
| Natural Environment | Carbon credit | digital | Carbon credit transaction |
| | Sand | physical | Sand supply chain and transaction |

(anybody can write) or private writing (only an authorized group of participants can write). Among the parties who can write, they may or may not have the same level of writing privilege.

Public writing are common for use cases targeting at general public. In the economic sector, peer-to-peer market of goods or services like the sharing economy for everyday tools [147] or for leftover foreign currency [148] are examples. Sharing of public research data in the educational sector is another example that is open for public writing [86], [87], [89].

In most use cases of our studied list, writing is restricted for authorized group of participants, either individuals or institutions. Private writing arrangement is more common because even use cases suitable for public writing might limit its participants, e.g., an e-commerce market could be open only for employees of a big company [135]. Many other cases are naturally fit for private writing. In the case of government sharing policy document with the public [61], the respective government agencies are the only ones authorized to write the respective documents to the blockchain. In a drug governance supply chain [119], government agencies, drug manufactures, wholesalers and hospitals, authorized patients are allowed to write drug related information to the blockchain. In an e-voting [64] or environmental monitoring [215] case, writers can be limited to citizens of the concerned region.

In addition to private versus public writing, it is helpful to understand different types of writing privileges writers may have. For example, in a peer-to-peer everyday tools sharing market [148], even though the lender and the renters have different roles, they are generally interchangeable because a participant can act as either a lender or renter when necessary.

But in a product ownership registration and tracking system [137], only the original manufacturers are allowed to register the new product they produce, and the rest of the public has instead the right to update the ownership attribute on the record of that product. These roles are not exchangeable and therefore the system involves asymmetric writing privileges.

## C. READERS

The readers are the parties that can view records on the blockchain database. Similar to the writing case, reading of records on the blockchain could also be public or private. Examples of public readability use cases are found in the governance, educational, economy and other sectors, for instance, government policy sharing [61], citizen participatory collaborative decision-making [5], [78], public research data sharing [86], [87], [89], sharing economy [147], [148].

Many blockchain use cases in various sectors enforce private reading where only authorized parties can read. For example, a certificate holder may show the certificate to specific employers or schools when requested [82], patients may release medical records for authorized personnels [110], supply chain status of goods may be viewed by designated partners [155].

## D. ASSET RECORD REQUIREMENTS - TRANSPARENCY, PRIVACY AND ANONYMITY

Transparency is a key value proposition of blockchain. However, use cases that involve off-chain assets often also require privacy and anonymity. The privacy and anonymity mechanisms seen in the studied use cases fall into three broad categories, PKI-based pseudo identity anonymity, content encryption, and dedicated privacy-preserving transaction mechanisms.

### 1) PKI-BASED PSEUDO IDENTITY ANONYMITY

The use cases often rely on the PKI public key based identities to provide pseudo-anonymity protection of the transaction parties, including identities of voters in e-voting [70], value-added-tax payers [73], event ticket holders [101], parties in the medical record sharing platform [109], [110], e-commerce customers [135], machine-to-machine electricity transactions parties [140], product ownership registration parties [137], peer-to-peer energy trading parties [186], [195], [206].

Traffic forensics and frequency analysis can yield patterns that compromise the anonymity in PKI public key based identity mechanism [110], [244], [245]. Therefore, additional measures are taken to improve the anonymity. For example, in an energy trading platform, the parties can generate new messaging addresses every time a new trade negotiation is initiated [197]. In an e-commerce rating system, in order to ensure that the feedback review cannot be linked back to the authoring customer, ratings are submitted only when there are enough other customers that could obfuscate the one actually submitting the review, and the system also enforces a time lapse between the transaction and the review submission [146].

### 2) ENCRYPTION FOR CONTENT PRIVACY

In the analyzed use cases, contents that are designated for a particular party are often encrypted by the party's public keys, providing privacy and allowing only the right recipient to see them. For example, the software licensing validation platform [92], [93] encrypts the licensing related data communicated using the end user's public key. The educational certificates platform [82] uses the certificate holder's public key for encryption. In an intelligent transportation system [165], vehicles crossing security domains encrypt the messages using the public key of the destination domain's security manager. In an anonymous messaging system for energy trading [197], a private person-to-person message is encrypted with the destination party's public key. Even though the message is broadcast and received by multiple parties, only the intended recipient can decrypt it.

Some of the surveyed systems also apply symmetric key encryption to preserve content privacy, such as records in a blockchain system that preserves author contributions on paper editing [88] and records in a supply chain use case [155].

### 3) DEDICATED PRIVACY-PRESERVING TRANSACTION MECHANISMS

Dedicated privacy-preserving transaction mechanisms have also been developed for blockchains and used by many use cases.

One promising solution is called zero-knowledge proof, which essentially allows a "prover" to prove that he has knowledge of a secret statement to a "verifier", without revealing the secret itself. When used in blockchain, it ensures that during the interaction, a verifier learns nothing about the transaction other than its validity. Therefore, the identity and amount of the transaction can be hidden from the nodes. The Zcash blockchain [246], rooted from Zerocash [247], implements zk-SNARKs (Zero-Knowledge Succinct Non-Interactive Arguments of Knowledge) and provides untraceable encryption that masks all transaction and allows only parties with the correct "key" to reveal the contents. However, the set of operations Zerocash allows is limited. Hawk [248] extends Zerocash's set of permitted operations to allow private transactions for arbitrary business logic. [102] is a blockchain-based lottery system that uses the Hawk model for privacy protection. In addition, the blockchain-based e-commerce reputation system in [146] proposed a NIZKs (non-interactive zero-knowledge proofs of knowledge) algorithm as the basis for preserving e-commerce reviewer's anonymity.

Permissioned blockchains can also offer their own integrated privacy mechanisms. For example, Hyperledger Fabric defines a collection of peer nodes as a logical channel, provides native per-channel based private transactions and a data collection mechanism to keep data private between participants of the same channel.

### E. UNDERLYING BLOCKCHAIN TECHNOLOGIES

The bottom part of our framework in Figure 1 is the blockchain infrastructure, highlighted by three of the key components: distributed consensus, smart contract and crypto token systems. We first take an overview look at the blockchain platforms used by the surveyed set of use cases, then examine these three respective components.

Among the 71 papers in the analyzed set, 43 of them declared specific blockchain types. Among them, the top four blockchains used are Ethereum (28), Hyperledger (9), Bitcoin (5) and MultiChain (5). While we can by no means claim this as an accurate quantitative assessment of the use of blockchains in research prototypes, it at least sheds some lights on the relative popularity of these well-known platforms among the research community.

### 1) ETHEREUM

Ethereum seems to be the most predominant platform. This is likely due to its status as the first established blockchain platform supporting full fledged smart contracts. It's use cases span almost all sectors we looked at, including voting [71], taxation [74], and collaborative urban decision-making [5], [78] in government sector; research paper authoring collaboration [90], digital media rights protection [102] and lottery system [102] in education, culture, science and innovation sector; clinical trial process improvement [105], medical data sharing [110], [114], supply chain of drug [121] and food [124] in well-being, health and safety sector; collaborative business process [133], [134], product provenance [137], automated sales systems [139], and sharing economies [147], [148] in the economy sector; sustainable transportation [172] in the transportation sector; energy efficiency and peer-to-peer energy market [178]–[180], [183], [184], [187], [195], and electronic vehicle charging [204] in the energy sector; and building operation management access control [214] in the built environment sector. It should be noted that while Ethereum is a permissionless blockchain, most of the research prototypes are run on the test net or a separate private Ethereum network due to their early stage nature.

### 2) HYPERLEDGER FABRIC

Hyperledger Fabric is a blockchain platform not only fully supports smart contracts but also provides built-in features particularly suitable for enterprise applications. It is one of the top two popular blockchain platforms we found in the literature set. It supports use cases including school information hub [243], event ticket system [101] in education, culture, science and innovation sector; medical data sharing [115], [116] and insurance process improvement [128]–[130] in well-being, health and safety sector; supply chain management [155] and vehicle-to-grid payment [203] in transportation sector. One reason the number of Hyperledger Fabric use cases is less than that of Ethereum in our surveyed literature is possibly because we have seen a lot of peer-to-peer market cases in our list, especially in the energy and economy sectors. Most of those cases used Ethereum since it has native crypto token payment support and is intended for the public, while Hyperledger Fabric does not integrate crypto token currency and is more for enterprise applications.

### 3) BITCOIN

The Bitcoin blockchain is commonly considered as only for cryptocurrency and not suitable for general purpose applications. This is mostly because its very limited scripting capability cannot support generic business logic. However, being the oldest and largest public blockchain network, it offers a secure and robust payment system that surpasses any other blockchains. So it is still one of the four popular blockchain platforms employed by researchers even in our study of non-cryptocurrency use cases. They appeared in publication from 2015 through 2018. The use cases include those that require limited business logic but place security as the foremost consideration, such as in e-voting [70], timestamping of research manuscripts [91], preservation of government documentation [61]. Other cases emphasize payment as the key proposition, such as electricity trading in micro grids [185], machine-to-machine micropayment [140], and electric vehicle charging payment [202].

### 4) MULTICHAIN

Interestingly, MultiChain [39] appears to be as popular as the Bitcoin platform found in our paper dataset. MultiChain is a permissioned blockchain. Compared to the other popular permissioned blockchain Hyperledger Fabric, MultiChain only provides very limited functionalities in implementing business logic, but it comes with its own crypto token currency. MultiChain was used by [73] to implement a value-added-tax system. The tax authority has control over who can join the blockchain and the system's crypto token is used to track tax amount. MultiChain also has another important feature called stream - which is suitable for recording time series key-value pairs. The stream feature, along with its other characteristics such as crypto token currency, makes MultiChain a frequent choice for those peer-to-peer markets that prefer a controlled set of participants. These applications often use the stream feature to publish offers and bids for the market, and the internal crypto token as currency for payment of asset transactions. Majority of the Multi-Chain use cases we found fall into this category. Among them the most popular ones are in energy domain, including electricity trading between producers and consumers in the chemical industry [189], or between devices on the smart grid to regulate supply and demand [182] and between household electricity prosumers [188]. In addition, Multi-Chain was chosen in the natural environment sector use case on trading carbon credits [218] and in a service trading market [132].

### 5) OTHER BLOCKCHAINS

A few other types of blockchains are also found in the research prototypes. Quorum [249] (which is based on an Ethereum core) is used by electronic health records sharing system [107] and vehicle life-cycle data sharing [151] because of its integrated support for private transactions. The ARK [250] blockchain was selected to record higher education course credits records [81] due to its flexibility in the number of programming languages its client implementations support. Reference [88] presents a system that measures research paper author contributions using the NEM [251] smart asset platform which is another blockchain that provide flexible business logic capability.

### F. CONSENSUS MECHANISMS

The consensus mechanisms are fundamental to the distributed database and often connected with the chosen blockchain platform. Among our studied dataset, the main types of consensus mechanisms used are proof-of-work, proof-of-stake and byzantine fault tolerance style ones.

### 1) PROOF-OF-WORK

Proof-of-work is the most adopted consensus mechanism, represented in at least half of the cases we studied. This includes all the 5 Bitcoin and 28 Ethereum blockchain use cases since proof-of-work are their default baked-in consensus mechanism. An additional 3 use cases also used proof-of-work even though the name of the specific blockchain is not mentioned, including for electricity meter reading security [173], an energy trading market involving industrial IoT and the grid [186], and peer-to-peer electricity trading between electric vehicles [206].

### 2) PROOF-OF-STAKE

In comparison, the number of cases using proof-of-stake is actually very small. A cross-cloud domain medical image sharing system [109] employs proof-of-stake to give more weights in the consensus process to providers who host more medical images. A peer-to-peer energy trading system in a micro grid [185] chooses proof-of-stake because it consume less energy than proof-of-work. The educational course credit recording system [81] uses a delegated proof-of-stake consensus system through the ARK blockchain platform it adopted.

It should be noted, however, Ethereum is in transition to proof-of-stake [252]. Many Ethereum-based use cases advocated moving away from proof-of-work to proof-of-stake, especially those energy trading blockchain use cases focusing on energy efficiency [179], [183]. Therefore, if we consider Ethereum to be proof-of-stake, the number of of proof-of-stake cases will be 31 and easily surpasses the remaining 8 proof-of-work cases.

### 3) BYZANTINE FAULT TOLERANCE

At least 17 of the use cases can be identified to support byzantine fault tolerance style consensus including the 9 Hyperledger Fabric cases. The 5 MultiChain-based use cases are also under this category because MultiChain's consensus are in spirit similar to practical byzantine fault tolerance. Another two use cases, one on micro grid peer-to-peer electricity trading [198] and the other on pharma supply chain tracing [118], use Tendermint [253] which is also based on byzantine fault tolerance. Quorum supports byzantine fault tolerance style consensus as well. Therefore the two Quorum-based projects for electronic health record sharing [107] and vehicle life-cycle data sharing [151] also belong to this category. Yet another work that used a byzantine fault tolerance based consensus is a customized blockchain for secure peer-to-peer sharing of documents among government agencies [63].

### 4) PROOF-OF-IMPORTANCE

The proof-of-importance consensus mechanism grants token mining privileges according to the user's importance. This importance is determined by taking into consideration not only the number of crypto tokens the user holds, but also the number of transactions a user made and with whom those transactions are made. Inclusion of transactions encourages the user to use the crypto tokens instead of merely holding them. A research manuscript editing record system [88] is based on the NEM blockchain platform that uses proof-of-importance consensus.

### G. SMART CONTRACTS

Smart contracts are instrumental in implementing the blockchain use case business logic. The original Bitcoin blockchain was not designed for smart contracts but is still capable of limited scripting functions. Newer blockchains such as Ethereum and Hyperledger Fabric provide full-fledged programming capability for all kinds of possibilities. In order to better understand the use of smart contract, it is helpful to categorize the use cases based on the business logic. However, we found that in most use cases the functionalities are intricately intertwined, making it extremely hard to come up with a reasonably small list of mutually exclusive categories. We therefore propose a taxonomy of a small number of functional models that simultaneously consider their interactions. They are short-named: ''Immutable Records, Access Control, Collective Decisions, and Peer-to-Peer Markets'', as shown in Figure 2. It should be noted that Immutable Records is the foundation category that applies to all other categories, while any of the other categories could intersect with each other as well.

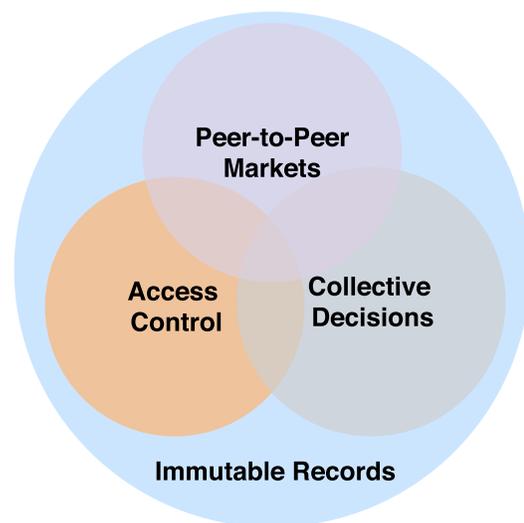

**FIGURE 2.** Business Model based Blockchain Use Case Classification.

### 1) IMMUTABLE RECORDS

Keeping immutable records is a fundamental utility provided by the blockchain technology. If the use case's chief value proposition is on data resiliency, smart contract functionalities may not be needed at all or could be achieved via very limited programming capabilities. In the studied set of use cases, we have seen these examples like preserving immutable records of public government documents [61], votes in e-voting [70] and timestamps of research manuscript submission [91].

### 2) ACCESS CONTROL

If the system needs to provide more advanced business logic in addition to tamper-resistant records, smart contracts become important. Access control is among the most common functionalities seen in the analyzed use cases. Access control is applicable to both the writers and readers and it controls who can write, what they can write, as well as who can read and what they can read. For example, in the education sector, student's educational certificates [82] and completed course credits [81] should be accessed by employers and other parties authorized by the respective students. School information about a region [243] can be accessed by related government agencies to facilitate educational resource planning. In the health domain, access control needs to be enforced for clinical trial agreements [105] and electronic medical records [107], [109], [110], [113]–[117]. Access control is also a common requirement in economy, transportation, and energy industries, such as for enrollment of manufacturers, claiming product ownership, recording product transfer from origin to post-supply chain [137], pharma supply chain compliance monitoring [121], tracing general goods supply chain record [155], monitoring smart meter readings [173], [175]. Even in the built environment sector, blockchain-based mechanism is proposed to enable decentralized access control among different building operating parties [214].

### 3) COLLECTIVE DECISIONS

Besides access control, generic business logic combining with participatory behavior can drive a new level of collective decision-making process. There are many examples seen in our studied set of use cases. A voting application [71] is able to assign voter eligibility, collect vote and manage the voting outcome. Insurance systems can monitor related asset and behavior status collectively to decide on the insurance policies and fulfill fine-grained insurance claims [128], [130]. In an energy domain application, [180] uses smart contract to optimize operation of the total available energy storage systems in order to achieve efficient use of shared resources and minimize external energy dependence. The governance domain also sees citizen participatory decision making use cases empowered by smart contracts. The system in [5] allows citizens to submit urban policy proposals, vote for urban planning decisions, select candidates to implement specific projects and reward them for performance. Reference [78] uses expert rules, open data and volunteer participants data to derive public health related collective decisions.

### 4) PEER-TO-PEER MARKETS

The original Bitcoin blockchain's primary utility is payment and digital asset exchange. Those functionalities are very popular in a large number of blockchain use cases. While basic form of blockchain payment does not require smart contracts, more sophisticated use cases for market-making applications may place access control for transaction parties, deploy specific market mechanisms, incorporate automated interaction with Oracles that will all benefit from various levels of smart contracts. For example, smart contracts can be used to implement a software development service trading market providing full functionalities including posting project requirements, submitting solutions, interacting with external Oracle for quality checking, and making payments [133]. Smart contracts can also enable a delivery service ordering system operated by autonomous unmanned aerial vehicle agents [134]. In peer-to-peer energy trading markets, smart contracts can be used to implement double auction market mechanism [184], enable automatic negotiation, settlement and payments [179], facilitate power flow estimation, optimization and control [178], [183].

### H. CRYPTO TOKEN SYSTEMS

The predominant crypto token system model in the use cases we studied is the utility model. An example of infrastructure level crypto token usage is [110], which is an Ethereum prototype that allows patients, doctors and authorized third parties to share medical data with permission management. The Ethereum's native Ether tokens are required for medical providers to perform their activities such as posting and updating records, accepting viewing permissions. Patients who wish to share their medical information also need to spend Ether or have the destination party fund them.

Use case specific application level utility tokens are very common. For example, we have seen crypto tokens used to represent tax credits in order to track actual amount of value-added-tax that should be imposed [73], to trace dividend paid in order to prevent frauds in dividend-based tax refunds [74], to transfer software entitlement [92], [93] and to document higher education course credits that students have completed [81]. There are also ways to reuse Bitcoin as application specific tokens, as in the colored coin method used by [185] for energy trading. It labels certain Bitcoins as issuance of energy token or transfer of energy token by setting their serial numbers to specify a special transaction type.

Payment is another popular token utility at the application level. Reference [140] is a system that enables Bitcoin-based micro-payment for a smart cable connected with a smart socket to pay for electricity. Reference [202] describes a system for Bitcoin payment between electric vehicles and the grid, leveraging the lightening network [254] scalability solution for Bitcoin blockchain. Other than using Bitcoin, there are also many use cases leveraging application-specific crypto token for payments, e.g., [134] presents a scenario of autonomous agents performing collaborative services which uses the blockchain application's internal crypto token for payments. Many energy trading markets also define their own crypto tokens for payments [184], [187]–[189], [197], [206].

Token incentives appear like payments, but they are also used to promote desired behaviors. Application level incentives are found in the studied use cases where they often play a crucial role in sustaining the business logic. For example, a product ownership management system for the post supply chain as in [137] can only be useful if sufficiently large number of users all register their products after every transaction. The system thus provides a crypto token incentive mechanism to encourage product owners' registration behavior. Similarly, the system in [172] use crypto token incentives to encourage urban citizens to engage in greener transportation methods such as cycling. In energy supply demand systems such as [183] and [179], crypto tokens are used to reward those parties who adhere to desired energy consumption profiles and help bring the system towards energy balance.

It is important to note that while some popular permissioned blockchain platforms like Hyperledger Fabric are not integrated with baked in crypto token payment systems, there could be use cases built on them that still require payment utilities. An example is the vehicle-to-grid payment system in [203]. It solves the payment problem by designing a specific type of transaction structure and keeping all historical transactions in the blockchain database for verification, thus effectively creating a payment utility in the system.

## VII. DISCUSSIONS

The results of our blockchain use case study, both the application-oriented review part, and the component-based analysis part, provide the basis for answering further questions in many dimensions, including the last three questions we posed in Section III-B.

### A. BLOCKCHAIN USE CASES AND URBAN SUSTAINABILITY GOALS

Our application-oriented review can provide us some initial evaluation on city sustainability targets. We look at four dimensions of sustainability as shown in Figure 3.

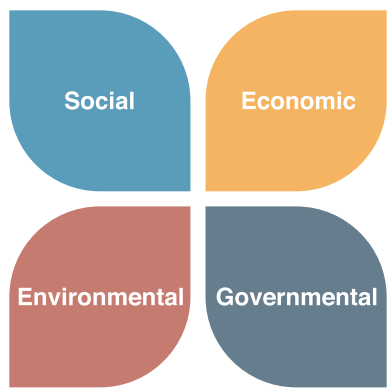

**FIGURE 3.** Four Dimensions of Sustainability.

Social, economic, and environmental constitute the triple bottom line on sustainability development introduced by the United Nations [255]. We can clearly see that all these three areas of sustainability are represented in various degrees in the use cases we surveyed.

The social domain focuses on people. In this respect, the well-being, health and safety area (Section V-C) is a heavily examined sector focusing especially on electronic health records, drug and food safety related use cases. Education and culture (Section V-B) is well represented mainly in improving the learning system and educational resource planning, as well as solving transparency problems in certain cultural entertainment programs. In contrast, the number of systems addressing the built environment (Section V-G) is relative few, and there is a lack of social diversity centered cases.

In the economic domain, we see a large of number of cases focusing on transforming commerce, sharing economy, collaborative business and other related aspects to improve prosperity (Section V-D). There are also efforts centered on enhancing the collaborative scientific research process to stimulate more innovations (Section V-B). Applications in transportation (Section V-E) such as the advancement of global goods supply chain and intelligent transportation systems, are bringing significant impact to both businesses and consumers in many other sectors of the society, potentially contributing to economic boost as well.

The environmental domain applications have concentrated on the green energy sector (Section V-F), manifested by numerous use cases concerning electricity prosumers, micro grid and electric vehicles, as well as peer-to-peer energy trading and supply demand balances. Air quality, carbon emission, sand mining have also been covered (Section V-I). However, there is in general fewer discussions on many topics in this domain such as on materials, water and land, waste management and climate resilience.

Last but not least, the governmental domain is special and has been considered a determine factor for the social, economic and environmental improvements [256]. It is also regarded as one of four pillars of sustainability [257], [258]. Our discussion on the governance and citizen engagement sector (Section V-A) illustrates that blockchain use in this domain is very much inline with and even provides crucial compliments to existing initiatives in meeting challenges for a truly citizen-empowered collaborative governance infrastructure.

It is worth pointing out that earlier work [7] shows existing smart city frameworks tend to focus significantly on social sustainability, reasonably on economic sustainability, but greatly under-represent environmental sustainability. In contrast, existing urban sustainability frameworks generally cover the environmental and social dimensions evenly, but almost ignores the aspect of economic sustainability.

Therefore, when we are considering blockchain initiatives for cities, it will be helpful to adopt a balanced perspective that incorporates all the social, economic, environmental as well as the governance aspects (in cases where the technologies are applicable), and we hope our preliminary assessments can help the establishment of a starting point on such efforts.

### B. BLOCKCHAIN USE CASE APPLICABILITY

#### 1) ASSESSMENTS WITH BLOCKCHAIN APPLICABILITY DECISION TREES

There exist well-known decision trees providing a list of questions about the assets, the writers and the readers to help evaluate the applicability of blockchain technologies for specific use cases [19], [20], [259]–[261]. They provide different levels of details but are inline at the core in determining whether and when a permissionless, permissioned or private blockchain should be chosen respectively.

The component-based analysis of our study provides an avenue to assess how the rules of these decision trees have been applied by the community. Interestingly, it is not hard to find reported use cases which may not be fully ''qualified'' for blockchain adoption or at least be inconsistent with the recommendations of some existing decision tree rules.

*Example 1:* According to the first criteria of [261], if the answer to "Are you trying to remove intermediaries or brokers" is no, then blockchain should not be used. In a use case for blockchain-based government decision preservation [61], government agencies create the documents and can directly share it with the public. So there is no third party intermediary involved and it seems not qualified for blockchain use. But in reality, this use case leverages the blockchain's immutable record capability as the key value proposition, not necessarily for removing intermediaries. A McKinsey report [262] also states that "Blockchain does not need to be a disintermediator to generate value".

*Example 2:* Based on [259], when other criteria of blockchain applicability pass and if the writers are all known but not trusted, a permissioned blockchain is recommended. In the same government document preservation use case [61], the writers are known as designated government agencies. If they are considered all trusted, then a blockchain should not be used. If they are not considered all trusted, then the rule would recommend a permissioned blockchain, but [61] uses the permissionless Bitcoin blockchain. This is possible presumably because appropriate off-chain mechanism can be used to recognize records written by the authorized parties. In this particular case, it could be based on government agencies' publicly released blockchain identity.

*Example 3:* In qualified blockchain use cases where the writers are not trusted and functionality control is needed, the recommendation would be to use a permissioned or private blockchain [261]. Then a product ownership tracking system [137] where many different writers require different levels of writing privileges would be more appropriate to use a permissioned blockchain. However, [137] uses the permissionless Ethereum blockchain and enforces the required control functionalities through its smart contracts capabilities. A similar point could be made on the reading side of the blockchain database. Specifically, it is natural for blockchain use cases with public readability requirements to adopt a permissionless blockchain. But for those use cases that require private readabilities, they do not necessarily entail a permissioned or private blockchain because they too may be implemented on a permissionless blockchain with appropriate smart contract mechanisms or may simply hide the information from public reading by encryption.

By highlighting the above inconsistencies between the reported use cases and the well-known decision rules, our goal is not to make judgements over right or wrong because both sides hold their merits. Instead we want to stress the significance for more systematic analysis, which we also attempted through our component-based framework review, to help reduce ambiguities in the terminologies and advance the overall knowledge in this infancy stage technology.

### 2) THE PHYSICAL-CYBER-CHAIN INTERFACE PROBLEM
While the three examples outlined in the previous Section VII-B.1 are relatively straightforward illustration of possible inconsistency between the rules and the actual cases.

There are many more subtleties when it comes to the topic of physical assets mapping.

According to the second decision rule in [261], if the answer to "Are you working with digital assets (versus physical assets)?" is no, blockchain should not be used. However, supply chain management systems deal with physical assets and yet they are among the most frequently reported use cases we found in the list of literature (e.g., [118]–[126], [136], [157]). In all these cases the physical assets are digitalized to be brought on-chain. But this physical and digital interface process has security risks. Reference [259] explicitly questioned the suitability of blockchain use in supply chain management unless the interface between the digital and physical world can be secured. In fact, this problem is not just for physical assets; it applies to all off-chain assets. As the assets part of our component-based analysis (Section VI-A) has shown, all the use cases on our list involve off-chain assets, which makes all of them vulnerable to this problem.

A closer look at the off-chain asset interface problem can reveal two sub-interfaces, both have security implications. One handles digitalization of physical assets (when applicable) which we call the physical-cyber interface; the other deals with actually placing the digital asset on-chain, which we call the cyber-chain interface. A security breach at the physical-cyber interface could be a tampered smart meter reporting a false electricity value, and a security violation at the cyber-chain interface could be a human knowingly or unintentionally uploading a wrong version of the digital document onto the chain. In both cases, the blockchain on its own is not able to detect the errors because those security problems happen off-chain. In other words, the blockchain only maintains an immutable record of whatever is committed on the chain, but it does not guarantee the correct association of on-chain and off-chain assets, or what happened to the assets before they were brought on-chain.

From an asset point of view, the only perfect asset for blockchains are indigenous on-chain assets that do not have to worry about the physical-cyber-chain interface. Those assets are commonly seen in the cryptocurrency space with the original Bitcoin being a great example. The original Bitcoins are minted on the chain, without tying to external assets, and have their entire history recorded on the blockchain. However, if the Bitcoins are used as a payment for some off-chain asset, they could still be associated with off-chain assets.

In certain circumstances, stakeholders can to some extent contribute to detecting off-chain asset mapping anomaly. For example, in a voting system [71], since the voters know what their respective votes are, they can check and ensure that the records of their own votes on the blockchain are consistent with what they intend to submit. Similarly, in an event ticket system [101], the ticket holders may be able to compare the ticket information they have and the actual ticket information record on the blockchain, therefore detecting mismatch between the two. In majority of other cases, however, stakeholder assistance can be very difficult, if not impossible.

For instance, if we are working with an electronic health records access control system [107] and somehow the health records provider submits wrong measurement value, or if we are dealing with an electricity system [173] and the meter device submits false data because of tampering, it is hard for the corresponding receiving party to notice the difference.

In summary, off-chain asset mapping and the physical-cyber-chain interface problem is universal in non cryptocurrency blockchain use cases. To deal with it, we should first try to avoid the interface whenever possible, e.g., the native digital form of the asset should always be preferred over a physical form (when applicable) in order to avoid the physical-cyber interface. For the cyber-chain interface that cannot be circumvented, we should carefully design the system to be resilient to possible risks. Stakeholder-assisted solution may apply to some cases, but most of the other cases require more sophisticated methodologies. It is worth noting that this off-chain asset mapping interface security issue is commonly treated as out of scope in the blockchain literature. We believe this is an important area that needs more substantive investigation for the justification of many blockchain applications.

## C. CROSS-SECTOR BLOCKCHAIN USE CASE CLASSIFICATION

An explicit goal of our work is to enable a review of the blockchain use cases with a horizontal perspective, i.e., compare use cases across different industries and potentially benefit from how they are used in distinct contexts to inspire more innovative and versatile solutions.

One way to facilitate cross-sector use case analysis is to classify them into certain common categories, regardless of their sectors. A universally accepted blockchain use case taxonomy does not exist, but related efforts are available. McKinsey stresses six categories of blockchain applications [262]: "Static Registry, Identity, Smart Contracts, Dynamic registry, Payments infrastructure and Other". This is helpful in the broad sense, though applying it to each specific use case is not always straightforward. For example, in its classification, land title, food safety and origin are considered static registry; while drug supply chain is considered dynamic registry. However, land title can be transferred and food can go through supply chain as well, so it is hard to draw the line between static and dynamic registry. We also found that smart contract is used or can be added in virtually all blockchain applications (only limited by the capability of the underlying blockchain infrastructure), so it is more natural to be considered a component of the system rather than a separate category by itself. Payment infrastructure is similarly a component that can be used by different types of blockchain applications. Even though the classification method does offer an "Other" category, it may defeat the purpose of classification if we place vast majority of use cases into "Others". In a related effort, Gartner highlights four types of blockchain applications [263]: "Record Keeper, Efficiency Play, Digital Asset Market and Blockchain Disruptor" (Digital Asset Market is also considered a special case of Blockchain Disruptor). This classification combines the role blockchain plays and the business model it enables.

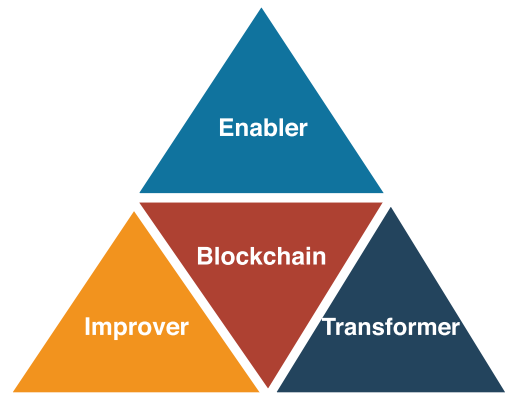

**FIGURE 4.** Role-based Blockchain Classification.

Inspired by these existing efforts, we propose two separate classification methods for blockchain use cases, one role-based and the other business model based.

For the role-based classification, we emphasize three broad types of roles that blockchain plays, as shown in Figure 4. First, the improver role is for those processes that are already conducted peer-to-peer without an intermediary. But use of blockchain makes the process more trustworthy and efficient. This is where we believe blockchain can create value without being a dis-intermediator and well answers the dilemma discussed in example 1 of Section VII-B.1. Second, the transformer role boosts process efficiency of existing intermediated processes by obsoleting the existing intermediary. Third, unlike the improver and transformer roles which are seen in existing business processes, the enabler role is found in newly emerged peer-to-peer business processes made possible by blockchains. It should be noted that merely stating a blockchain usage area does not allow one to deduce the particular role that blockchain plays in that use case. Different ways of blockchain usage may be applied to the same context and result in improvement, transformation or enablement. As an example, we can consider a blockchain use case that manage car life-cycle information similar to [151]. If we just use blockchain to keep an immutable record of some car attributes like ownership or maintenance, that is an improver case; if we use the blockchain to conduct used car buying and selling transactions directly between two peer parties, that becomes a transformer case because it removes the transaction intermediary in the traditional process; if we further extend the blockchain use to enable innovative insurance or other services, that could make it an enabler case.

Since business logic in blockchain applications are determined by smart contracts, we have introduced our business model based classification earlier when we analyzed smart contracts usage of the use cases in Figure 2 of Section VI-G. The four intersected categories, "Immutable Records, Access Control, Collective Decisions, and Peer-to-Peer Markets" are meant to stimulate insights across sectors. In particular, when we design a specific solution within a given context, our references do not have to be only prior use cases in the same sector, but could be those with the same business model category from very different sectors. Let us take the Access Control category (Section VI-G.2) as an example. Use cases in this category span entirely different sectors in education, economy, healthcare, transportation, energy, built environment and so on. While common access control mechanisms define a list of absolute permissions for entities, a built environment use case [214] adopts an interesting and different approach. It uses delegated permissions system which relates the permissions of entities among each other, through a graph of "delegation of trust". The result is a system in which all entities with a permission on a resource are equally capable of delegating their permissions to other entities. It can be imagined that this type of mechanisms can be applied to other scenarios that require dynamic and equal delegation of access, and those scenarios can be in sectors totally different from the built environment. It is this type of cross-sector knowledge sharing that we hope our work could help cultivate.

## VIII. CONCLUSIONS AND FUTURE WORK

Blockchain is potentially a disruptive force in the next wave of urban development initiatives, along with decades of sustainable and smart cities efforts. Many cities around the world have already started their race for the blockchain future through regulatory actions and comprehensive pilot projects in both public and private domains. However, there is great concern about the infancy stage of the blockchain technology and the paucity of understanding on how it can be applicable to future cities. This work is an effort to narrow this gap by leveraging the collective knowledge from use cases reported by the scientific research community.

### A. CONTRIBUTIONS

Our main contributions are summarized below:

1) Following a systematic literature review methodology, we first examined 159 blockchain literature from academic journals and conference proceedings that cover concrete use cases and systems. They were structured and discussed within 9 industrial sectors that are well recognized as essential to sustainable and smart cities. We found that some of the sectors, like natural environment, water and waste management, built environment in general receive less attention than other sectors such as energy, transportation, economy, healthcare, education, and governance. Regardless of the sector, there is a common list of challenges that blockchain applications face, such as infrastructure performance and scalability, standardization and interoperability, asset data security and privacy, smart contract security, as well as legal and regulatory issues.

2) We proposed a component-based analysis framework to facilitate a common understanding for blockchain

use cases. A subset of 71 papers were studied regarding their associated assets, writers, readers, and the underlying blockchain infrastructure with the distributed consensus algorithms, smart contracts and crypto tokens systems. Among the surveyed literature, we found Ethereum and Hyperledger Fabric are not surprisingly the top two most frequently used blockchain platforms, with Ethereum having an edge especially on many peer-to-peer market-based applications. It is interesting to see both Bitcoin and MultiChain are also among the top four platforms used. On the distributed consensus aspect, proof-of-work and byzantine fault tolerance are the two most used mechanisms. But proof-of-stake usage could easily surpass proof-of-work if Ethereum finished its transition to proof-of-stake. In addition, the utility crypto token model is widely used in many applications for representing application-specific assets, making payments and giving incentives.

3) The implication of blockchain use cases towards the urban sustainability goals was discussed. Our application-oriented use case review demonstrates that all four pillars of urban sustainability: social, economic, environmental and governmental are represented in the surveyed literature. In the governmental domain, blockchain-empowered citizen participatory collaborative urban governance model is even considered the exact answer to overcome key problems of existing solutions. Meanwhile, more efforts on certain areas of the environmental domain could contribute to a more balanced blockchain treatment on sustainability.

4) We investigated the relationship between well-known blockchain applicability decision trees and actual system prototypes reported by the research community. Specifically, we explained the inconsistencies found between the two and highlighted why component-based system analysis like ours can be beneficial. In addition, we discussed the physical-cyber-chain problem that is suffered by all the use cases surveyed and advocated its importance.

5) To facilitate cross-sector analysis, we offered two methods for classifying blockchain use cases. The role-based approach groups use cases into blockchain as "Improver, Transformer and Enabler". The business model based approach delineates three intersected categories: "Access Control, Collective Decisions, Peer-to-peer Markets", and the fourth category called "Immutable Records" that is the foundation to all the other three. We elaborated how these taxonomies compare with existing ones from the industry, and illustrated how they can help bring cross-sector insights for blockchain use case analysis.

### B. LIMITATIONS

The contributions of this paper need to be considered in light of its limitations. Due to the enormous amount of literature on blockchain, we have to confine our scope of review only to papers focusing on concrete use cases, with sufficient system level coverage and in the explicitly specified sectors. We also acknowledge that the manual screening process of filtering thousands of papers down to under 200 and the sector placement of each use case inevitably introduces subjectivity. As a result, even high quality papers could have been excluded. Yet to the extent possible, we tried our best to select the sufficiently comprehensive and relevant set of papers to support our analysis.

## C. FUTURE WORK

There are a number of directions this work can be taken further. First, for the application-oriented use case review, it will be interesting to assemble more blockchain use cases that are actually in operation in the industry (some of them have been reported in the literature, but others might not). This would allow a comparison between the use cases in operation and those at early research prototype stage, and help identify how the research and industry could benefit from each other through a tighter interaction. Second, the current connection of our application level review with the sustainable and smart cities frameworks is at the macro sector level. Future work can look into the more micro level associations between blockchain use cases and specific urban sustainability and smart city framework indicators. These analysis can then pave way to a possible future standardized assessment framework of "blockchain for cities" which can become guiding principles for urban policy makers. Third, another way to extend this work is to dive deeper vertically into the elaborated use cases: for example, leveraging both the component-based framework and the classification mechanisms to come up with lists of system design references at the component or more finer-grained level for each type of use cases in any sector, maximizing the knowledge sharing for blockchain industrial professionals among all disciplines.

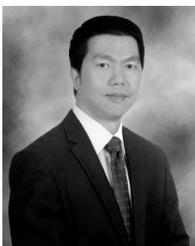

**CHARLES SHEN** received the B.S. degree in electrical engineering from Zhejiang University in 1997, the M.Eng. degree in electrical and computer engineering from the National University of Singapore in 2001, the Ph.D. degree in electrical engineering from Columbia University in 2010, and the Executive M.B.A. degree from the Columbia Business School in 2017.

He is a Research Scientist in civil engineering and engineering mechanics and the Co-Director of the Advanced Construction and Information Technology Laboratory, Columbia University. Prior to his current position, he was a Senior Member of Technical Staff at AT&T. Before AT&T, he had conducted research for various periods at the Department of Computer Science, Columbia University, IBM T. J. Watson Research Center, Telcordia Technologies (formerly Bell Communications Research, now part of Ericsson), Samsung Advanced Institute of Technology, and Singapore's Institute for InfoComm Research (A*STAR).

Dr. Shen's current research interest is in the application of innovative information technologies, such as blockchain and artificial intelligence in sustainable urbanization domains. His past experience includes extensive research on mobile computer networks and applications, such as scalability of IP telecommunications networks, architecture, security and privacy of cloud-based mobile Internet of Things services. His extended work in the architecture, engineering, and construction industry covers topics, including integration of Internet of Things with building information modeling, social impact project finance, and public-private-partnership. His interdisciplinary research outcome has yielded eight awarded U.S. patents, three IETF RFC Internet standards and specifications, and dozens of peer-reviewed papers at journals and conferences sponsored by ACM, IEEE, IFIP, and ASCE.

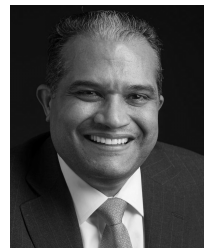

**FENIOSKY PENA-MORA** received the M.S. and Sc.D. degrees in civil engineering from the Massachusetts Institute of Technology in 1991 and 1994, respectively. He is the Edwin Howard Armstrong Professor of civil engineering and engineering mechanics, a Professor of earth and environmental engineering, and a Professor of computer science at Columbia University. He also directs the Center for Buildings, Infrastructure and Public Space at Columbia.

From 2014 to 2017, he was on a public service leave serving in the role of Commissioner of the New York City Department of Design and Construction (DDC). There, he was responsible for over 1200 projects valued in excess of U.S. $15 billion, undertaken by more than 1400 workers and 1320 consultants. Under his leadership, more than 860 construction projects, valued at more than U.S. $9 billion, started or completed; the agency received more than 80 design and professional awards; the agency also committed more than U.S. $5.4 billion in new contracts by improving the capital project procurement process, each one of these accomplishments a record for DDC. Prior to his public service leave at DDC, he was the Dean of the Fu Foundation School of Engineering and Applied Sciences and the Morris A. and Alma Schapiro Professor of engineering at Columbia University. In this post, he was responsible for setting the school strategic direction and managing its operation and growth to over U.S. $400 million endowment, U.S. $200 million annual operating budget, 4500 students, and 400 staff and faculty members. Previously, he was an Associate Provost and the Edward William and Jane Marr Gutgsell Endowed Professor at the University of Illinois at Urbana-Champaign. Before joining the faculty at Illinois, he was the Gilbert W. Winslow Career Development Professor at the Massachusetts Institute of Technology.

Dr. Pena-Mora has authored or co-authored more than 220 scholarly publications. He holds six patents and one provisional patent. He is a fellow of the Chartered Institute of Buildings as well as an Elected Member of the Dominican Republic Academy of Sciences and the United States National Academy of Construction. He was a recipient of multiple awards, including the Presidential Early Career Award for Scientists and Engineers (known as PECASE), the National Science Foundation CAREER Award, the Walter L. Huber Civil Engineering Research Prize of the American Society of Civil Engineers, the ASCE Computing in Civil Engineering Award, and the ASCE Construction Management Award.

• • •